\documentclass[10pt, conference, letterpaper]{IEEEtran}

\usepackage{graphics}
\usepackage{verbatim}
\usepackage{epsfig}
\usepackage{latexsym}
\usepackage{graphicx}
\usepackage{balance}
\usepackage{algorithm}
\usepackage{algorithmic}
\usepackage{graphicx}
\usepackage{array}
\usepackage{bm}
\usepackage{color}
\usepackage{amsmath}
\usepackage{enumitem}
\usepackage{lipsum}
\usepackage{epstopdf}
\usepackage{makecell}
\def\BibTeX{{\rm B\kern-.05em{\sc i\kern-.025em b}\kern-.08em
		T\kern-.1667em\lower.7ex\hbox{E}\kern-.125emX}}

\begin{document}

	\title{Preventing Handheld Phone Distraction for Drivers by Sensing the Gripping Hand}
	
	\author{
		\IEEEauthorblockN{Ruxin Wang, Long Huang, Chen Wang}
		\IEEEauthorblockA{School of Electrical Engineering and Computer Science,
			Louisiana State University, Baton Rouge, LA 70820 USA}
		\IEEEauthorblockA{Email: rwang31@lsu.edu, lhuan45@lsu.edu, chenwang1@lsu.edu}
	}

    \maketitle
	
	\begin{abstract}
Handheld phone distraction is the leading cause of traffic accidents. However, few efforts have been devoted to detecting when the phone distraction happens, which is a critical input for taking immediate safety measures. This work proposes a phone-use monitoring system, which detects the start of the driver's handheld phone use and eliminates the distraction at once. Specifically, the proposed system emits periodic ultrasonic pulses to sense if the phone is being held in hand or placed on support surfaces (e.g., seat and cup holder) by capturing the unique signal interference resulted from the contact object's damping, reflection and refraction. We derive the short-time Fourier transform from the microphone data to describe such impacts and develop a CNN-based binary classifier to discriminate the phone use between the handheld and the handsfree status. Additionally, we design an adaptive window-based filter to correct the classification errors and identify each handheld phone distraction instance, including its start, end, and duration. Extensive experiments with fourteen people, three phones and two car models show that our system achieves 99\% accuracy of recognizing handheld phone-use instances and \textcolor{black}{0.76-second} median error to estimate the distraction's start time.
\end{abstract}
	
	\sloppy
	\section{Introduction}
\label{sec:intro}

Using a handheld device while driving is a dangerous behavior. The driver can be impacted by all three types of distractions from the phone (i.e., visual, manual, and cognitive), which increases the risk of crashing by up to 23 times~\cite{lee2013fatal}. Though law enforcement and insurance penalty policies help raise public awareness and lower car accidents, they achieve limited effects. Reports show that handheld device distractions cause 1.6 million crashes annually in the U.S., and 2018 alone, over $400,000$ people were injured or killed in car accidents related to cell phone use~\cite{thezebra2021texting}. Since the COVID-19 pandemic, a 17\% increase in driver phone use has been found, because more people attempt to take Zoom calls, read Instagram messages, or text while driving~\cite{zendrive2021zendrive}. More efforts are still in urgent need to reduce the driver's handheld phone use to improve traffic safety.

There has been active work on using the smartphone itself to prevent distractions. By recognizing when the phone user is driving, the smartphone could automatically turn on the do-not-disturb mode and prohibit phone use (e.g., delaying messages and routing calls to voice mail). For example, the cellphone handovers and signal strength variations can be used to recognize a phone in a moving car~\cite{gundlegaard2009handover}. 
To further discriminate the phone user to be a driver or passenger, researchers have developed the in-vehicle localization methods, which estimate whether the phone is closer to the driver seat or the passenger seat~\cite{yang2011detecting,wang2015determining}. However, most users refuse to disable phone services completely while driving though acknowledging the dangers~\cite{cbs2014why}. They may have concerns about missing important notifications and calls during long-distance driving. They may also prefer to use the less distracting and legally allowable handsfree phone operation, ask the passenger to read/reply or pull over to a safe area to cope with emergencies. Thus, it is more practical and effective to prevent a driver from reaching out to the phone rather than disabling all phone services for an entire trip. More specifically, we need to know when a driver holds the phone to stop the phone distraction at once. 

This work aims to capture the precise timing (e.g., start, end and duration) of each distracted driving instance, which is a critical input to numerous safety systems for taking immediate safety measures. For example, knowing when the driver picks up the phone, all Apps could be shut down by the phone at once except the emergency calls. And the nearby automobiles (especially smart cars) could be notified to take precautionary measures. Additionally, such information could be used for determining who is at fault in a car accident or personalizing insurance rates. The prior work to monitor the driver's phone use mainly relies on monitoring the display on/off, the phone lock status, the phone lifting action~\cite{li2016dangerous}, 
and the phone dynamics related to distracting phone activities (e.g., texting and calling)~\cite{ahmed2018leveraging}. But based on such indirect phone-use indicators, these methods are hard to determine the detailed timing of each distracted driving instance. Moreover, they have limited abilities to cover the diverse phone distraction scenarios and are not sufficiently reliable in the practical in-vehicle environment, which is noisy.  

Different from the prior work, we propose to monitor the driver's phone use between the handheld and the handsfree status by directly sensing the gripping hand, which enables eliminating the distraction or reducing the impact at once. We develop a phone-use monitoring system based on acoustic sensing, which starts to work when the phone user has been identified as the driver by existing methods~\cite{yang2011detecting,wang2015determining}. The smartphone's speaker emits ultrasonic pulses periodically to sample the phone-use status. The acoustic signals traveling on the device surface could be uniquely damped, reflected and refracted by a gripping hand. The resulting signals reaching the smartphone microphones are then different from the scenarios when the phone is placed on a seat, cup holder, pocket or phone mount, due to their unique materials and contact areas. Based on that, our system accurately detects when the driver grabs/holds/drops the phone within a second. Because no additional hardware is required, users of our system can continue to use their existing cars without technological restrictions.

In particular, we develop a CNN-based distracted driving prevention system, which continuously monitors the phone-use status and captures distracted driving activities promptly. Each pulse sound received by the smartphone microphone is used to derive the short-time Fourier transform, which describes the unique time-frequency characteristics of the signal interference caused by the gripping hand or a support surface in the vehicle. Moreover, we develop a CNN-based binary classifier to discriminate whether the smartphone is handheld or handsfree. The binary classification decisions from the periodic sensing pulses thus provide the continuous monitoring of the phone use. We further develop an adaptive window-based filter to correct the error samples, whose phone-use status toggles too quickly between handheld and handsfree to represent a human activity. Based on that, we determine the start, end and duration of each handheld phone-use activity.

\textbf{Our contributions can be summarized as:} 
\begin{itemize}[leftmargin = *]
\item This work proposes a continuous phone-use monitoring system to eliminate the driver's handheld device distraction. The proposed system leverages active acoustic sensing to detect when the phone is held by the driver's hand and take safety-enhancing measures immediately. 

\item We derive the short-time Fourier transform from the sensing sound to describe the phone's contact object and develop a CNN-based algorithm to discriminate the handheld phone use from various handsfree scenarios. 

\item We design error correction mechanisms to process the phone-use sampling results and facilitate capturing each complete handheld phone-use instance and its detailed timing information accurately in noisy in-vehicle environments. 

\item Extensive experiments in the practical driving environment show that our system captures the complete handheld phone-use activities and accurately determines their start times.

\end{itemize}

	\section{Related Work}
\label{sec:related}
 

There has been a rising interest in monitoring unsafe driving behaviors. The vehicle's speed, acceleration and deflection angle can be estimated from the smartphone sensor data to recognize the dangerous driving behaviors~\cite{eren2012estimating,fazeen2012safe}. To improve the drivers’ awareness of their driving habits, Chen \textit{et al.} further classify the abnormal driving behaviors among different vehicle maneuver types by using smartphone sensors~\cite{chen2015d}. Xu \textit{et al.} focus more on the driver's attention and use Doppler shifts of the phone audio signals to sense the inattentive driving events, such as eating, drinking, and turning back~\cite{xu2017leveraging}. But none of these works could effectively address the handheld device distraction, one of the leading causes of traffic accidents. 

The existing research efforts to prevent handheld phone distraction are on discriminating the phone user to be the driver or the passenger based on its in-vehicle location. 
Yang \textit{et al.} propose a relative-ranging system, which sends acoustic signals in a programmed sequence from the stereo car speakers and captures the time differences of their arrivals at the phone to determine whether it is closer to the driver seat or the passenger seat~\cite{yang2011detecting}. Wang \textit{et al.} use the phone's inertial sensors to measure its centripetal acceleration when the vehicle makes turns. By comparing to a reference point, they estimate whether the phone is on the right or left side of the car~\cite{wang2015determining}. Chu \textit{et al.} release the requirement of additional infrastructure and rely entirely on the smartphone sensors to differentiate the micro-activities between the driver and the passenger, such as with which foot to enter the car first and along which direction to fasten the seat belt~\cite{chu2014smartphone}. There are also infrastructure-free methods to recognize the phone user during driving, which localize the phone based on its motion dynamics or camera views~\cite{paruchuri2015detecting,wahlstrom2019smartphone}. However, these methods are far from satisfactory to address the handheld device distraction, as they cannot detect when the distracted driving happens to take proper safety measures right away, which requires capturing the interaction between the phone and the driver.


There are several solutions to capture phone-driver interactions based on cameras. For example, 
Chuang \textit{et al.} monitor the driver's gaze direction by using the smartphone front camera~\cite{chuang2014estimating}. A recent work installs multiple cameras in the car to capture the interaction between the driver and the phone, which complements the blind spots of each single camera~\cite{jin2020deep}. However, these vision-based methods are limited by light conditions (especially at night), camera view angles or high installation overhead. 

We propose to monitor phone-driver interactions based on sensing the gripping hand. There have been several studies on detecting grips of mobile devices. For example, the smartphone’s rotations, vibrations and touch events can be measured by inertial sensors and the touchscreen to infer the user’s phone-use postures, such as with which hand (or both hands) to hold the device and which finger (e.g., index finger and left/right thumb) to operate on the screen~\cite{goel2012gripsense, park2015study}. These are motion-driven approaches. Ono \textit{et al.} attach a pair of vibration motor and receiver on the phone case to recognize the user’s hand postures ~\cite{ono2013touch}, and Kim \textit{et al.} achieve similar functions based on acoustic sensing~\cite{kim2019smartgrip}. Both methods use a support vector machine as the classifier. However, the above studies all assume the phone is already in the user’s hand and then recognize the type of phone grip. Few of them investigate distinguishing a handheld phone from that placed on many other surfaces such as a table, seat, and phone mount. Furthermore, it is unknown whether they could work in the in-vehicle environment, which suffers from complex acoustic noises and vibration noises related to the engine, road conditions, and the wind. More importantly, none of them is able to demonstrate the user-phone interaction monitoring and capture the phone-grip start, end and duration.

	\section{Background and System Architecture}

\subsection{Distracted Driving Instance}
This work aims to prevent the distracted driving caused by handheld phone use. We define a \textit{distracted driving instance} as the handheld phone-use activity, which begins from the driver's hand reaching the phone and ends until the phone is dropped off. This entire period is subject to the combination of all three types of distractions (i.e., visual, manual and cognitive). Compared to the single-distraction-type activities, such as checking the navigation system (visual), making a handsfree phone call (cognitive) and eating/driving (manual), the handheld phone use is the most dangerous and is prohibited by law. Therefore, to prevent the handheld phone use of a driver, one efficient and direct way is to detect when and how long the driver's hand holds the phone and then disable or restore the phone services accordingly. It also facilitates sending early warnings, notifying the nearby automobiles, assisting the law enforcement, and personalizing insurance rates.

\subsection{Sensing the Gripping Hand Acoustically}
\label{subsec:feasibility}

\begin{figure}[t]
	\begin{center}
		\includegraphics[width=2.228in]{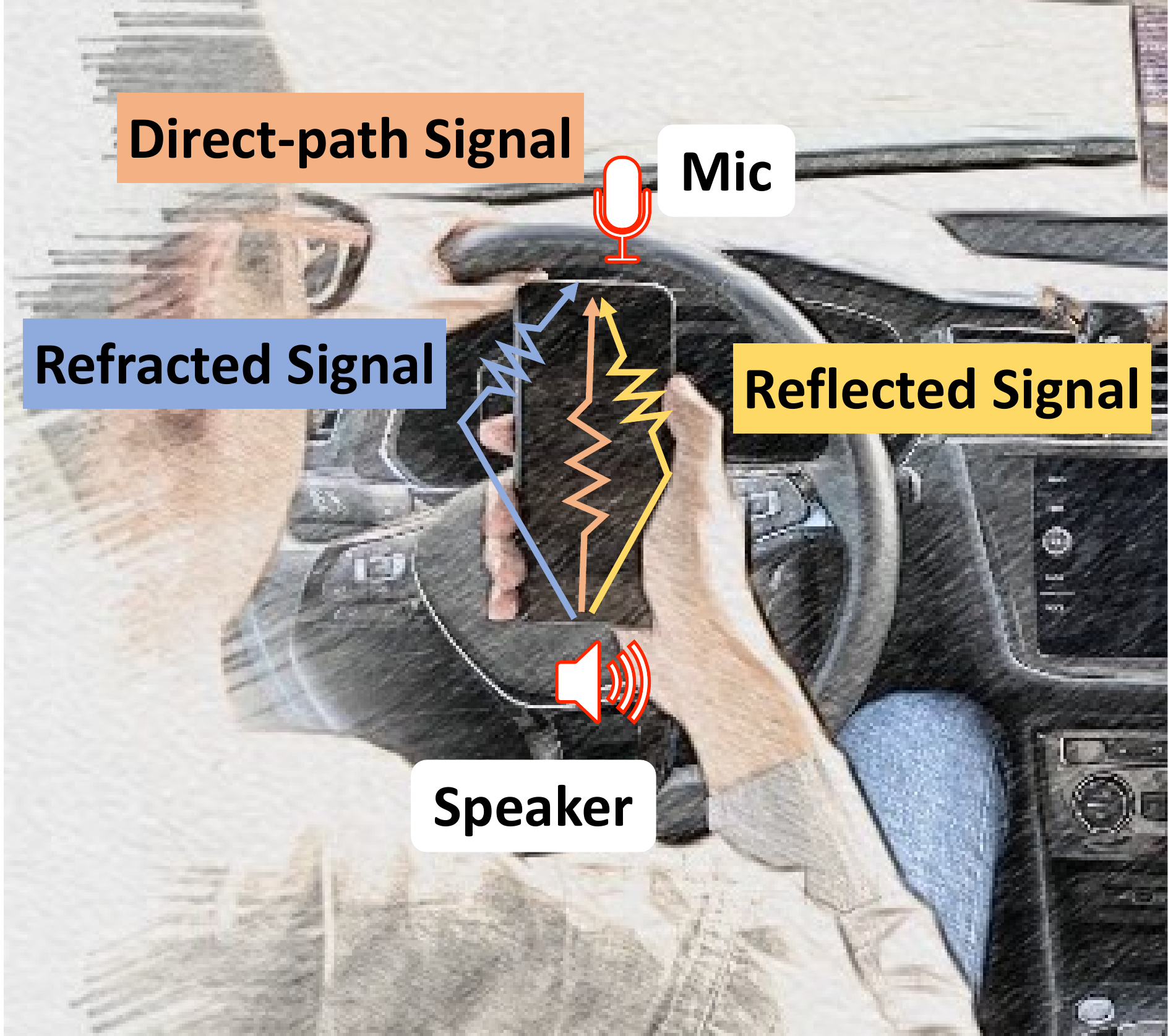}
	\end{center}
	\vspace{-2mm}
	\caption{Illustration of the acoustic signal interacting with the driver's hand. }
	\vspace{-4mm}
	\label{fig:scenario_plot}
\end{figure}

We leverage the acoustic signals that propagate on or near the smartphone surface to sense the gripping hand or other objects that come in contact with the phone. In particular, we use the smartphone speaker to periodically send ultrasonic signals for sensing. The signal traveling on the phone case would be interfered by the driver's gripping hand or the support surfaces on which the phone is placed, such as the seat and center console. The resulted sound reaching the smartphone microphone contains the useful information that could describe how differently the original signal is damped, reflected and refracted by the gripping hand and the support surfaces. Figure~\ref{fig:scenario_plot} illustrates how the acoustic signal interacts with the driver's hand, where the sound recorded by the microphone includes the damped direct-path signal, the reflected signals and the air-borne refracted signals (near-surface). These signal components are mainly determined by the material, area and pressure of the contact surface. Because the hand's skin, geometry and gripping strength are distinct from that of any support surface in the car, the gripping hand can be discriminated by acoustic sensing. 

\begin{figure}[t]
	\begin{center}
		\begin{tabular}{ccc}
			\includegraphics[width=0.29\columnwidth]{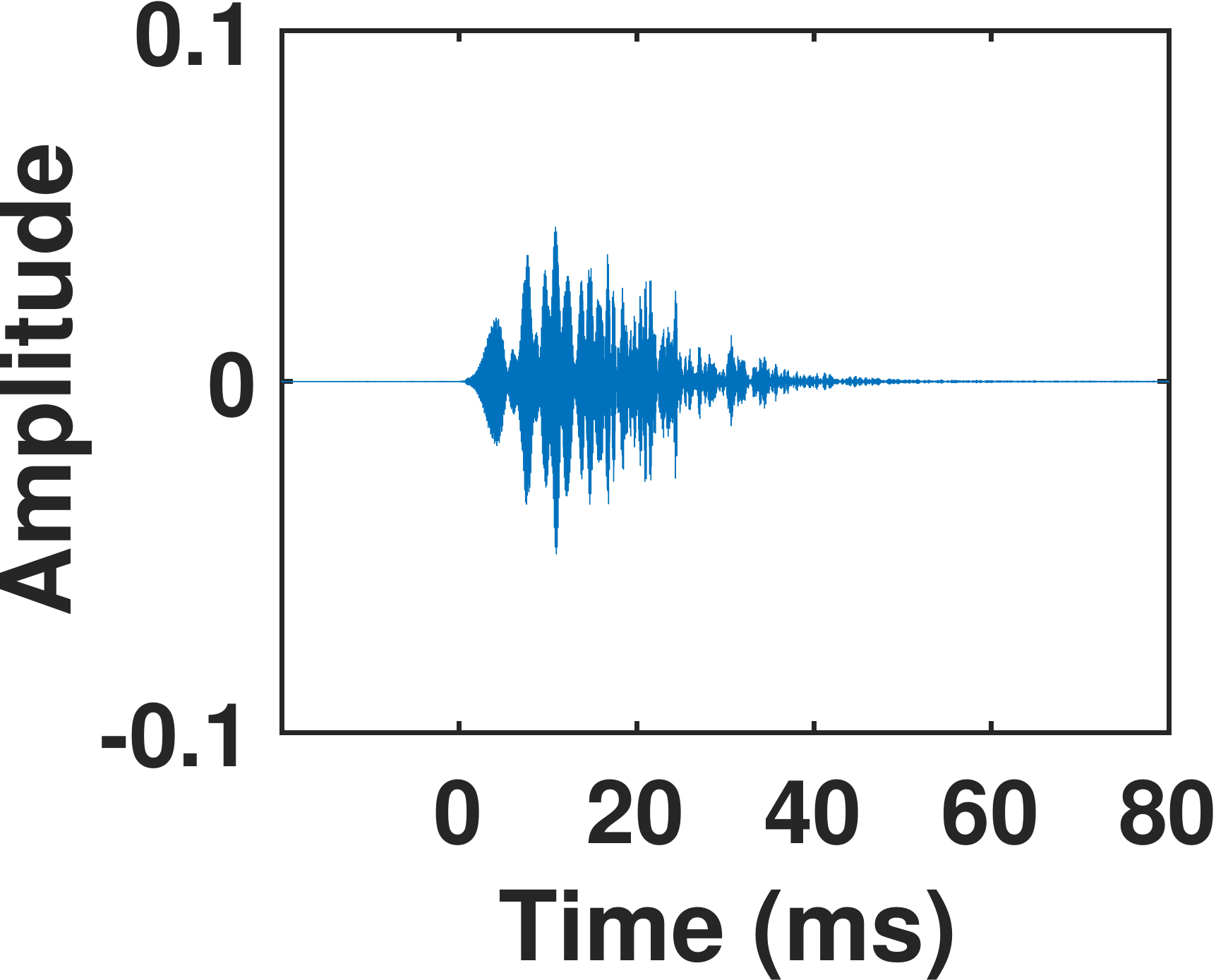}
			&
			\includegraphics[width=0.29\columnwidth]{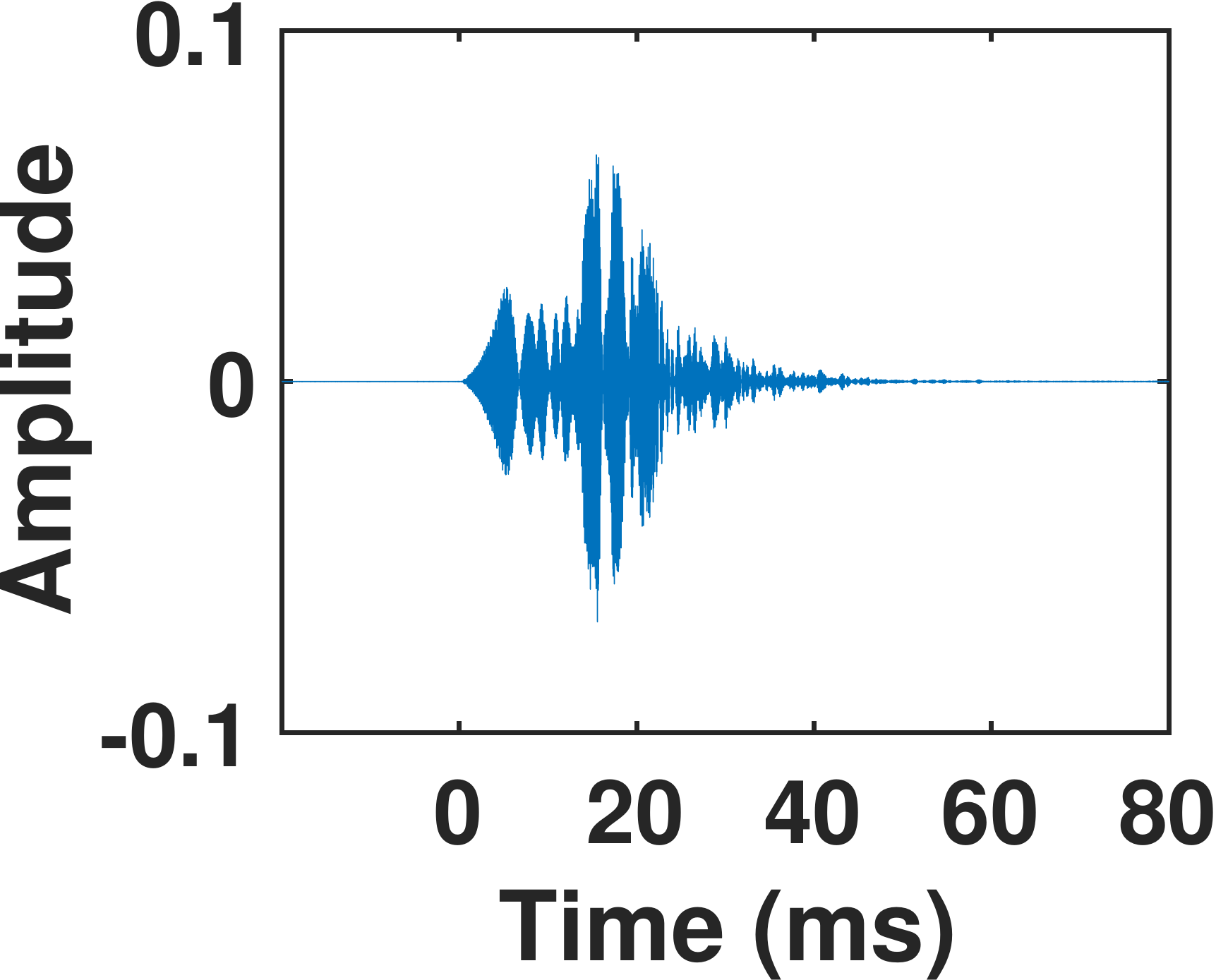}
			&
			\includegraphics[width=0.29\columnwidth]{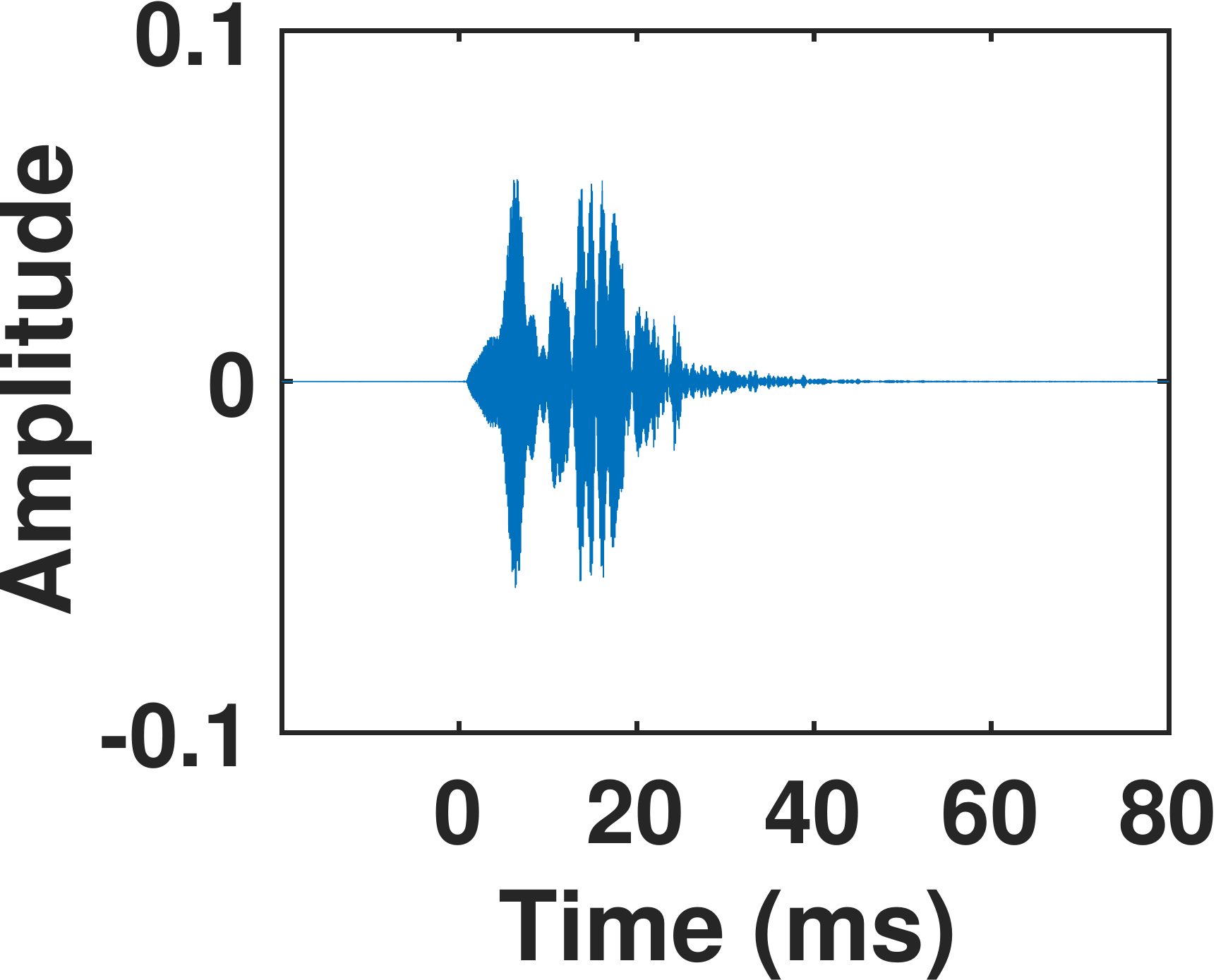}
			\\{\scriptsize(a) Handheld} & {\scriptsize(b) Center console} & {\scriptsize(c) Cup holder} \\
			\includegraphics[width=0.29\columnwidth]{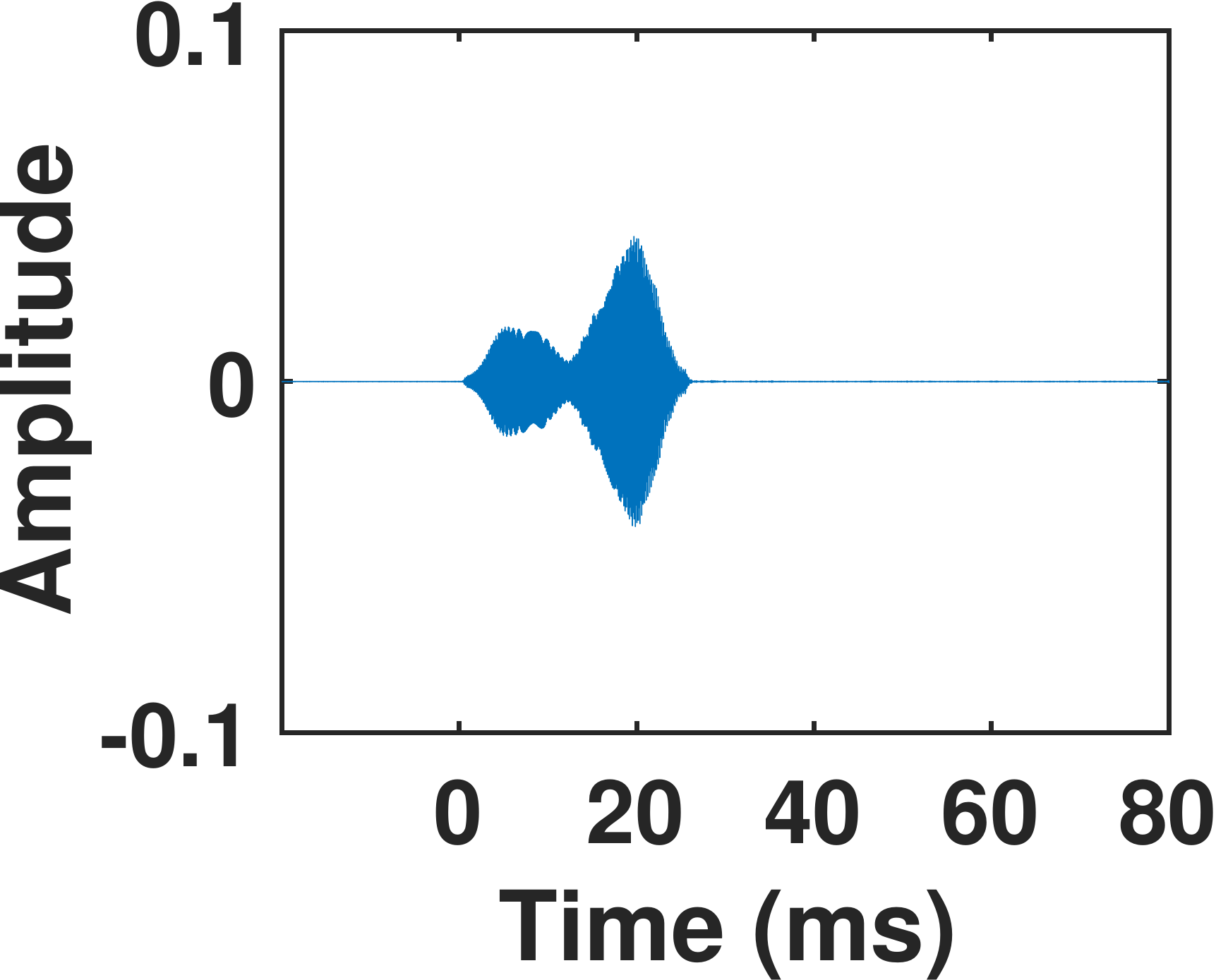}
			&
			\includegraphics[width=0.29\columnwidth]{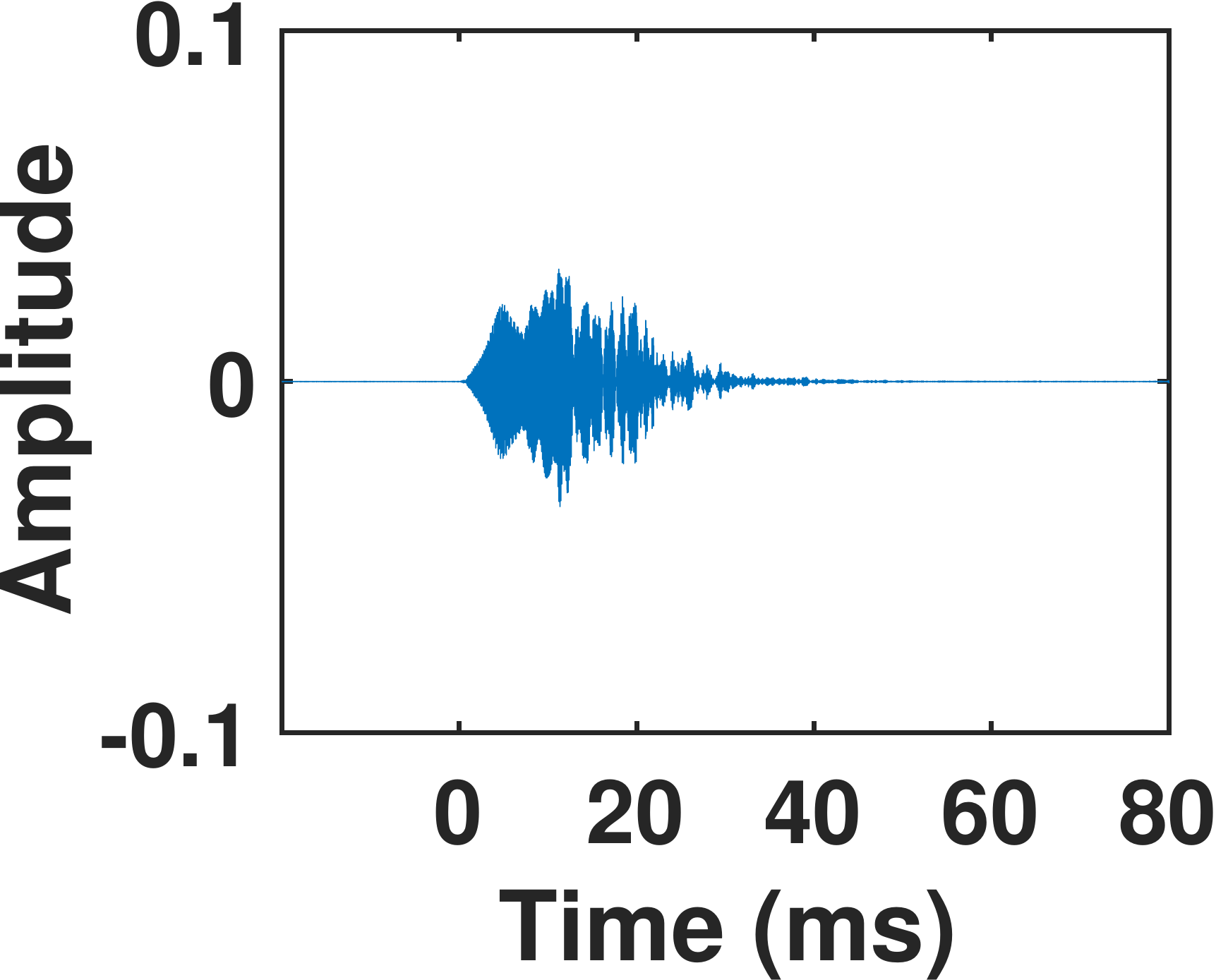}
			&
			\includegraphics[width=0.29\columnwidth]{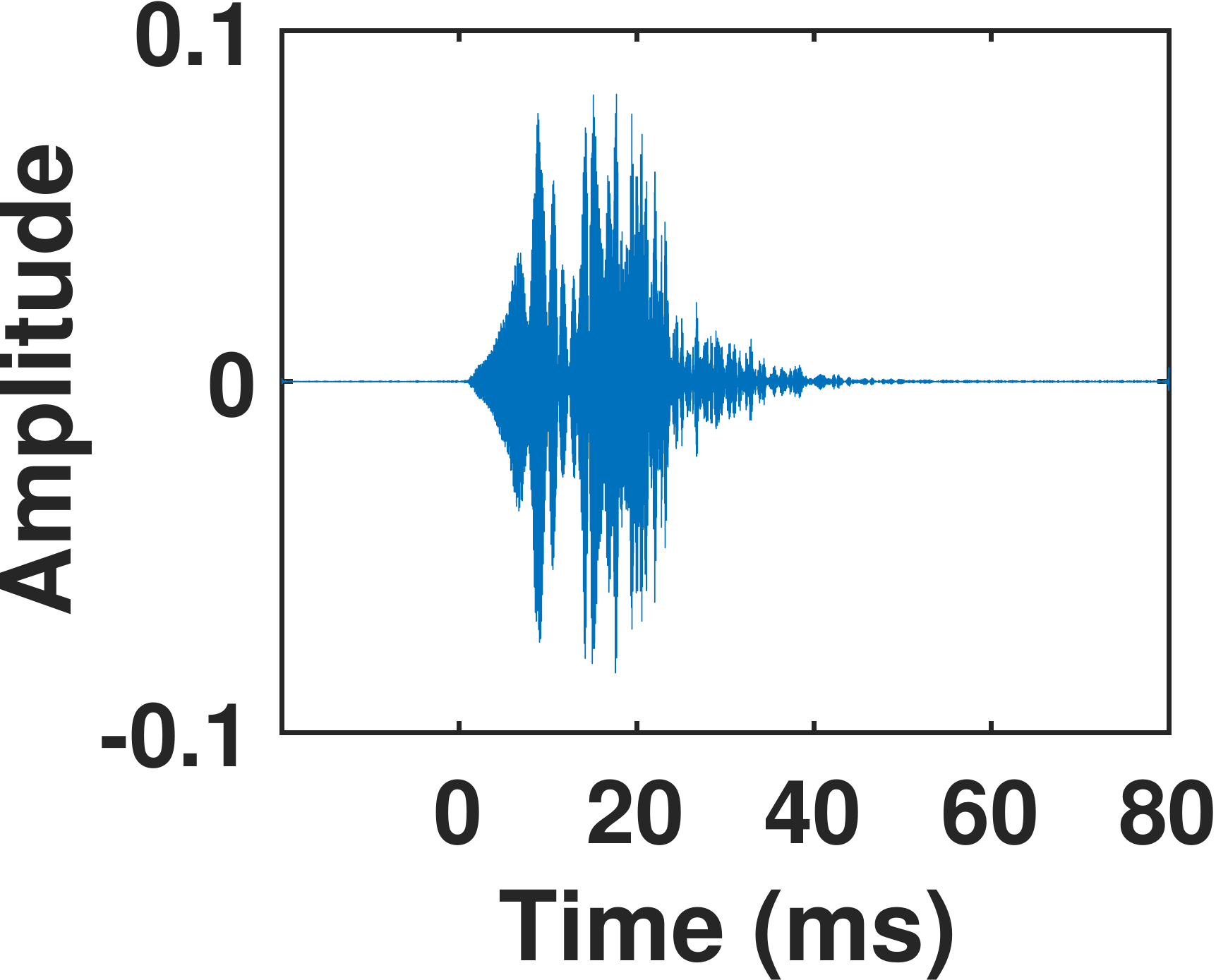}
			\\{\scriptsize(d) Pocket} & {\scriptsize(e) Seat} & {\scriptsize(f) Smartphone mount} 
		\end{tabular}
	\end{center}
	\vspace{-1.8mm}
	\caption{Acoustic response of different smartphone placement. }
	\vspace{-5mm}
	\label{fig:placement_waveform}
\end{figure}

To show the feasibility, we play an ultrasonic chirp sound using the smartphone's own speaker, which sweeps from 18kHz to 22kHz in 25 ms. Figure \ref{fig:placement_waveform} shows the waveforms of the recorded sounds when the phone is on six different support surfaces in a car, including a hand. We observe that the microphone recorded sound is distinguishable in the waveforms among all the six phone placement scenarios, which shows the potential of differentiating the gripping hand from the other phone placement scenarios. Moreover, while the sensing signal sweeps along the frequency, its amplitude is reinforced or suppressed with different scales, and at the same frequency, the amplitude change is also unique for each support surface. This phenomenon reflects the frequency diversity of the sound to sense the various support surfaces, which motivates us to use the sound with rich frequencies rather than a single frequency to achieve robust sensing. 

\subsection{Challenges}
\label{subsec:challenge}
We also face some challenges when using acoustic signals to sense the gripping hand. Specifically, we find that the microphone keeps receiving sounds for a long time after the sensing signal stops at 25ms as shown in Figure~\ref{fig:placement_waveform}. These sounds are mainly the environmental reflections, which are much stronger in the confined space of the vehicle compared to indoor or outdoor scenarios. They also heavily rely on the in-vehicle phone locations and should not be used for analysis. One exception is the in-pocket scenario, because the fabric of the pocket is a good sound-absorbing material, which damps the outward sounds and reduces the echoed back sounds significantly. As a result, the in-pocket waveform is more different from that of the other five scenarios. In comparison, the handheld scenario is harder to be differentiated from the center console, cup holder, and phone mount scenarios. We thus rely on deep learning to recognize the handheld scenario.

Furthermore, different people's hands may exhibit slight differences when holding the phone, due to the individually unique hand shapes and gripping strengths. Even the same person may hold the device slightly different when texting, scrolling and calling. These variances need to be considered and addressed. Furthermore, our acoustic system must be able to work under the noisy in-vehicle environment, where the background noises are resulted from the different road conditions, driving speeds and car audio sounds. Additionally, to accurately estimate the start and end of a distracted driving instance, our system needs to handle the classification errors and the noisy transient states when the phone is being grabbed or dropped off.

\begin{figure}[t]
	\begin{center}
		\includegraphics[width=3.3in]{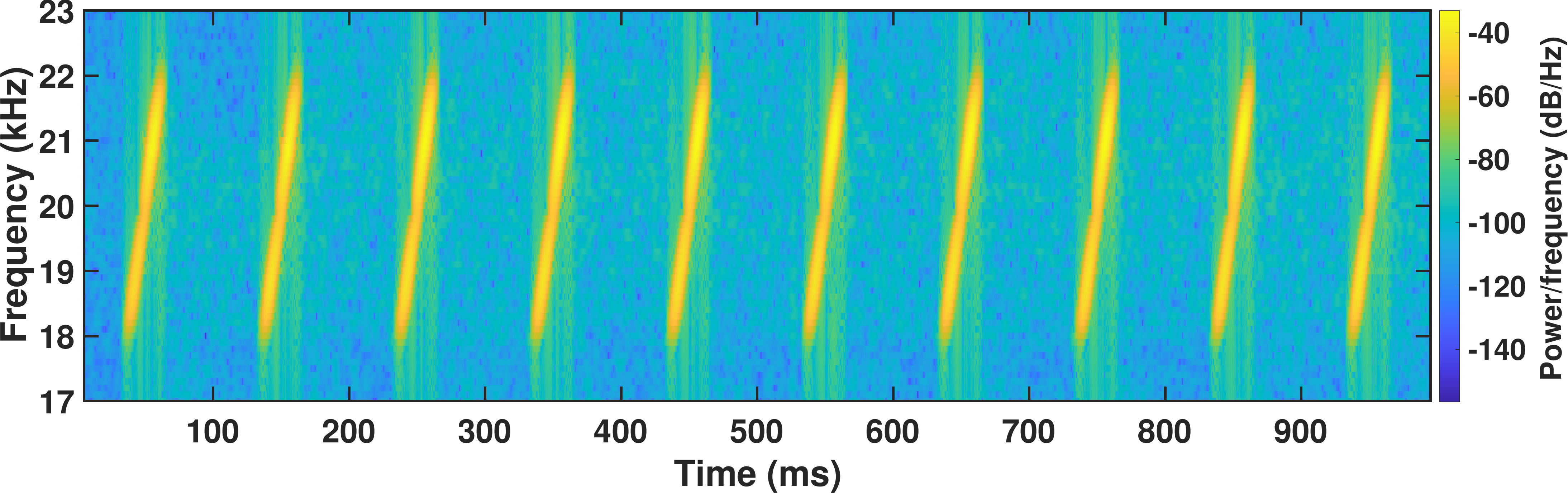}
	\end{center}
	\vspace{-3mm}
	\caption{The periodic ultrasonic pulse signals for sensing. }
	\vspace{-5mm}
	\label{fig:stimulus}
\end{figure}

\subsection{System Design}
The goal of our work is to eliminate the handheld phone-use distraction based on detecting the gripping hand. To achieve the goal and address the above challenges, we develop a phone-use monitoring system, which sends unique signals for sensing and uses a deep learning-based algorithm to recognize the various in-vehicle phone-use statuses. Our system can work with existing phone localization methods~\cite{yang2011detecting,wang2015determining} to more effectively eliminate the handheld phone distraction. For example, our system could start after the phone user is identified to be the driver. Alternatively, our system could continuously sense the phone use status, and once it is in a hand, the phone localization method further confirms if this is the driver's hand. 

\subsubsection{Sensing Signal Design}
The sensing signal is used to interact with the object that is in contact with the smartphone and capture its characteristics in the acoustic domain to discriminate whether the phone is in the driver's hand or on a support surface of the vehicle. Based on our feasibility study and challenge analysis in Section~\ref{subsec:feasibility} and~\ref{subsec:challenge}, we design the periodic ultrasonic pulse signal as shown in Figure \ref{fig:stimulus}. In particular, each pulse signal lasts for a short period (i.e., 25ms), and every two pulses are separated by a stop period (i.e., 75ms). The short pulses suffer less from the echo sounds which usually last for much longer, and the stop time reduces the interference between adjacent pulses. Only the 25ms pulse sound is used for analysis. 

Moreover, each pulse signal is designed to sweep from 18kHz to 22kHz to leverage the rich frequency information, which facilitates capturing more characteristics of the object in contact with the phone. Besides, this high-frequency range is not impacted much by the in-vehicle noises, which are mainly on lower frequencies. The sounds in these frequencies are also demonstrated to be hardly audible and not invasive~\cite{ashihara2007hearingthresh}. Furthermore, to reduce the spectral leakages and the speaker hardware noises caused by the sudden frequency jumps at the start and the end of each pulse signal, we apply a Hamming window to smooth the two ends of each pulse. As a result, the pulse signals could sample the phone-use status ten times per second to capture the gripping hand.

\subsubsection{System Flow} 

\begin{figure}[t]
	\begin{center}
		\includegraphics[width=2.45in]{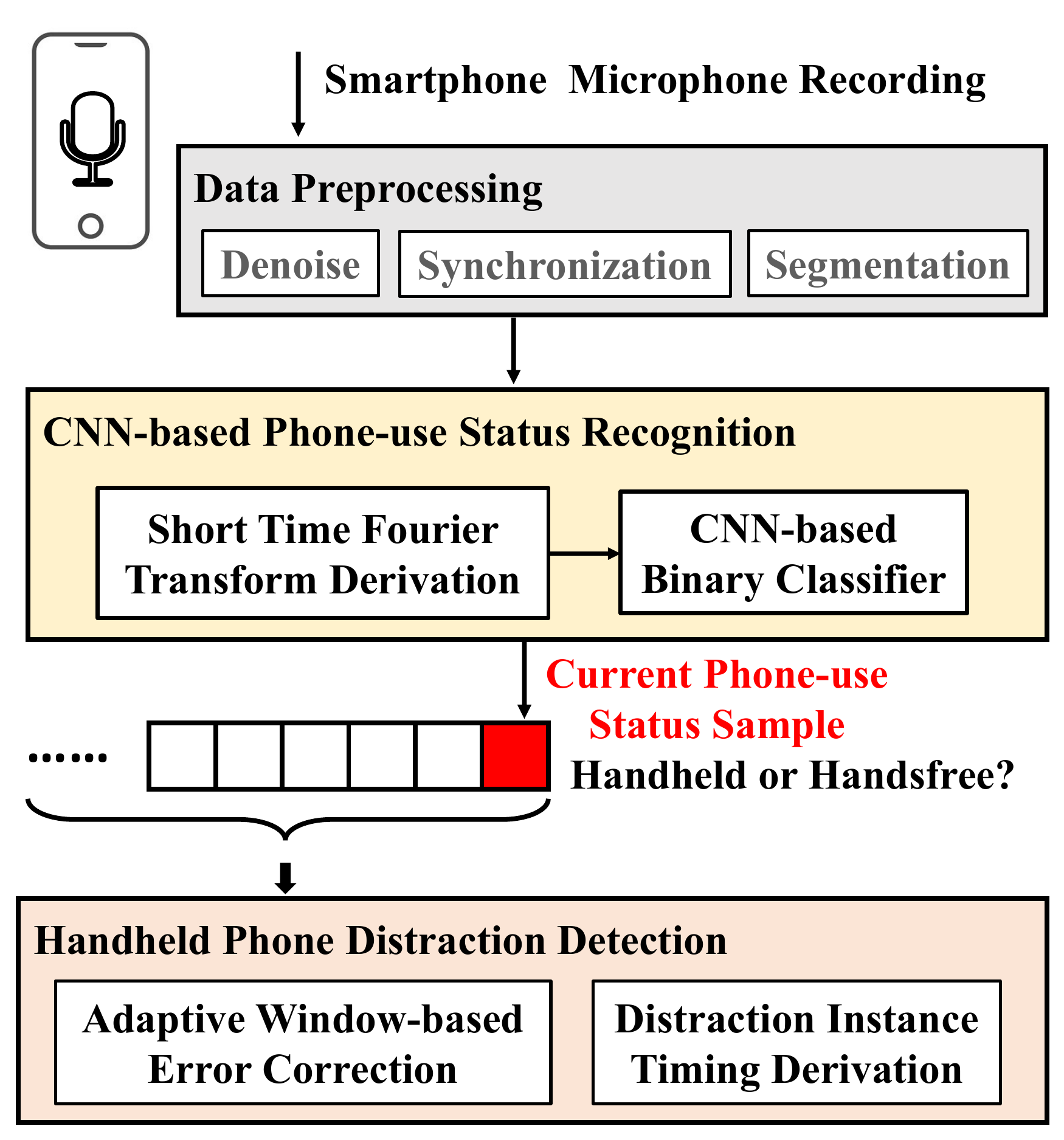}
	\end{center}
		\vspace{-3mm}
	\caption{The architecture of our system. }
	\vspace{-5mm}
	\label{fig:system_plot}
\end{figure}

The architecture of our system is shown in Figure \ref{fig:system_plot}, which takes the smartphone microphone recording as the input. \textcolor{black}{It is important to note that our system can be integrated with voice assistant to reuse its recording without incurring additional recording tasks.} The \textit{Data Preprocessing} is performed first to prepare the data for analysis. It applies a bandpass filter to remove the noises outside of the sensing signal frequency range and synchronizes the data by referring to the original audio. Based on that, we can find the start and end of the pulse signal to obtain the pulse segment, which contains the information about the phone-use status.  

The core of our system consists of two components. The \textit{CNN-based Phone-use Status Recognition} processes the pulse sound and recognizes the phone-use status at the current sampling point. Based on a series of the most recent phone-use status samples, the \textit{Handheld Phone Distraction Detection} further captures the distracted driving instance and determines its timing. Specifically, we derive the Short-Time Fourier Transform (STFT) from the pulse sound to describe the time-frequency characteristics of the contact object in the acoustic domain. The 2D STFT is input to the CNN-based binary classifier to discriminate the phone-use status between handheld and handsfree. Furthermore, we develop an adaptive window-based error correction method to further examine the current status sample and correct the classification errors based on the results of the recent pulse sounds. We then use a threshold-based method to determine whether a distracted driving activity occurs. If the phone-use status toggles back and forth too quickly, it is unlikely to be from human action and is corrected. Once a distracted driving instance is confirmed, the system would take safety measures immediately, such as disabling phone services, sending early warnings, and notify nearby automobiles to be aware. If the phone is detected to be dropped off, the phone services can be restored. By capturing the phone-grabbing and drop-off time, we obtain the detailed timing information of each distracted driving instance.

	\section{Method Design}
\label{sec:acoustic}
\subsection{Data Pre-processing}

After obtaining the data from the microphone buffer, we first pre-process it for denoising, synchronization and segmentation. In particular, we design a bandpass filter with the 18kHz-to-22kHz passband to reduce the noises outside of the sensing signal's frequency range. For example, the engine, road and wind noises can be removed, which are mainly on the frequencies below 6kHz \cite{cerrato2009automotive}, and the car audio sound impact could be reduced. After denoising, we can focus better on the sensing signal changes caused by different contact objects. 

Next, we run a synchronization scheme to precisely locate the pulse signal in the microphone data. Specifically, we iteratively shift the microphone data $\hat x$ and compute its cross-correlation with the original pulse signal $x$. The shift length leading to the maximum cross-correlation coefficient indicates the time delay between the two signals as expressed by
\begin{equation}
	\label{eq:find_delay}
	\setlength{\abovedisplayskip}{3pt}
	\setlength{\belowdisplayskip}{3pt}
	delay = \underset{m}{argmax}\sum^{N-m-1}_{n=0} \hat x (n+m) x(n),
\end{equation}
where $m$ is the number of samples to shift. After subtracting this delay, we can find the start and end of the sensing sound by referring to the original pulse signal. The resulted 25ms pulse segment is used for further analysis. We further normalize the amplitude of the pulse segment to be within the range $[-1,1]$. It is important to note that the pulse signal is generated every 100ms, and the 75ms microphone audio that comes after the pulse are mainly the echo sounds. This audio part is heavily affected by the phone's in-vehicle location and is discarded. 

Furthermore, most smartphones are embedded with two microphones for noise cancellation (e.g., one at the top and one at the bottom). By using the two acoustic channels, we can leverage the spatial diversity to capture more characteristics of the contact object. Therefore, we use the two mics to independently sense the contact object and integrate their results to make a decision, which reduces the errors of each single mic and is robust.

\subsection{Short-Time Fourier Transform}

\begin{figure}[t]
	\begin{center}
		\begin{tabular}{ccc}
			\includegraphics[width=0.29\columnwidth]{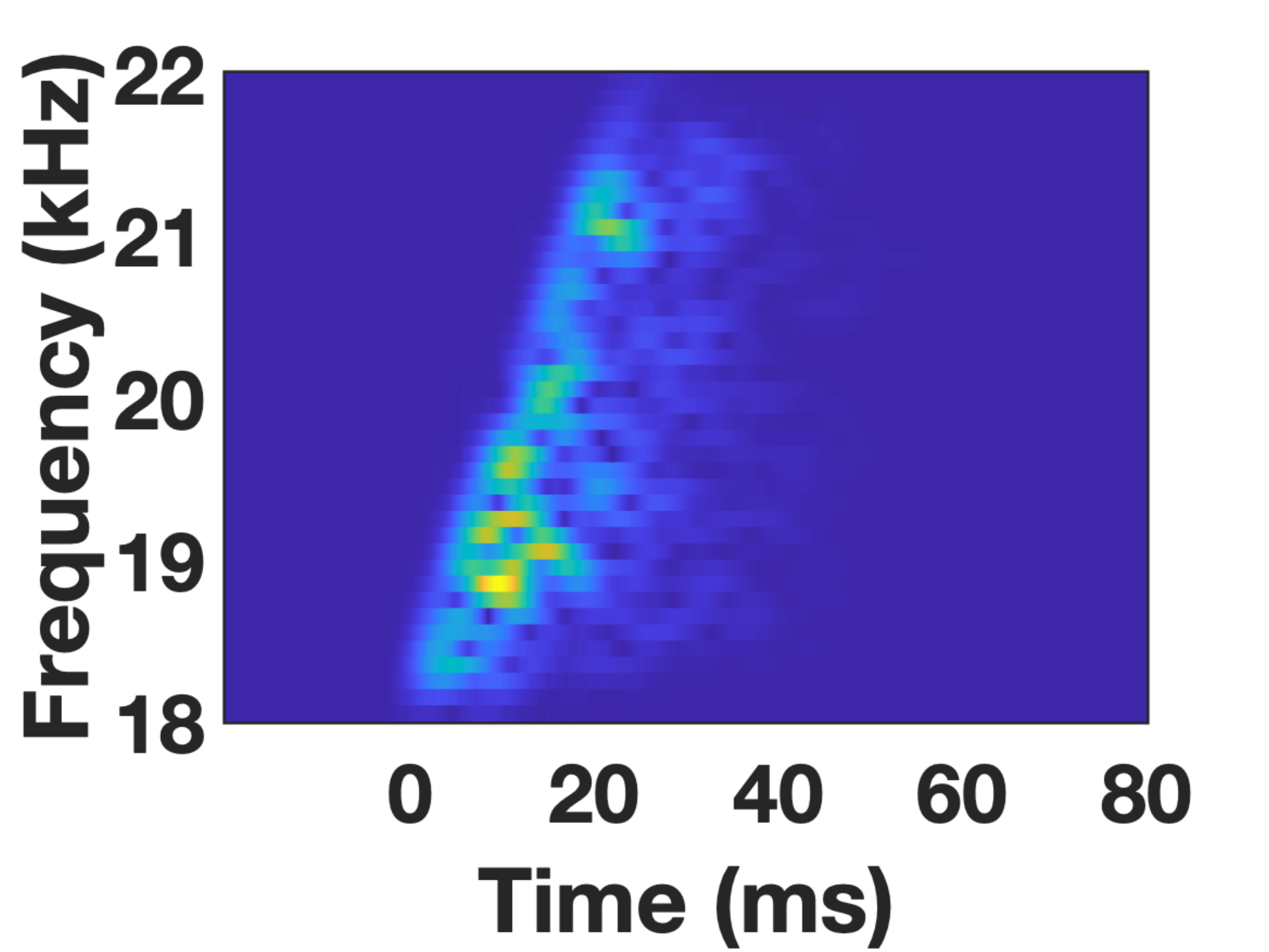}
			&
			\includegraphics[width=0.29\columnwidth]{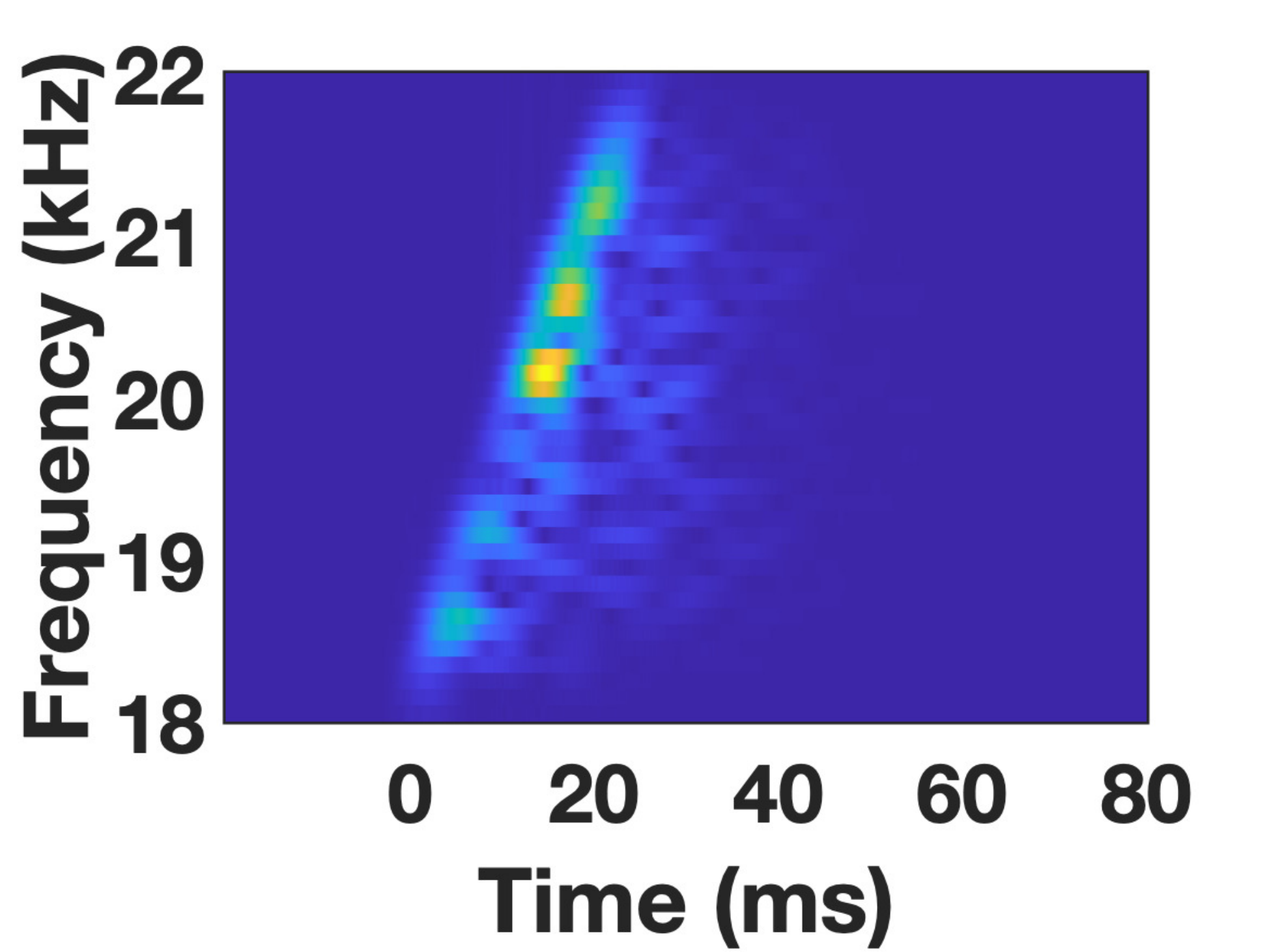}
			&
			\includegraphics[width=0.29\columnwidth]{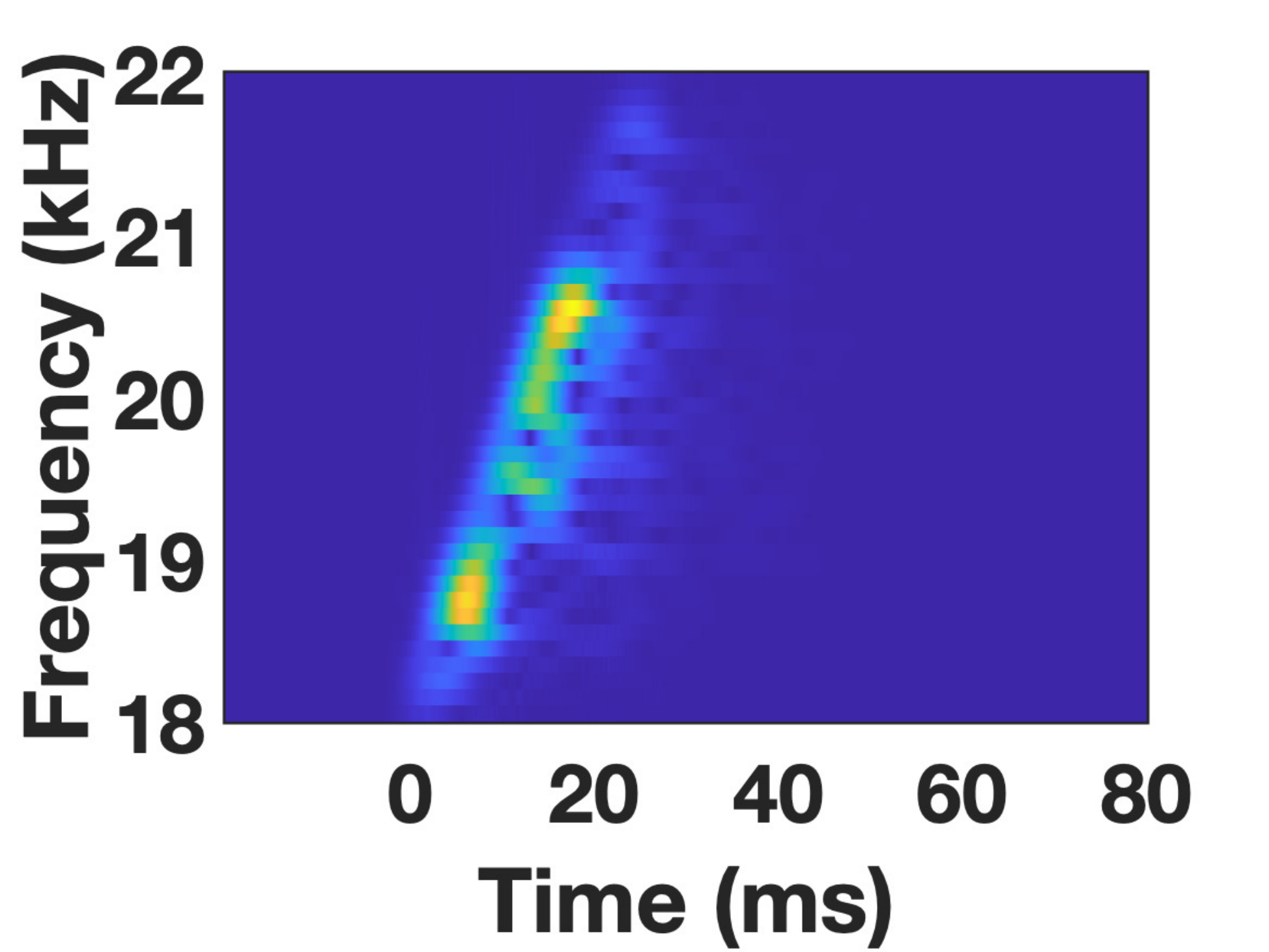}
			\\{\scriptsize(a) In hand} & {\scriptsize(b) On center console} & {\scriptsize(c) On cup holder} \\
			\includegraphics[width=0.29\columnwidth]{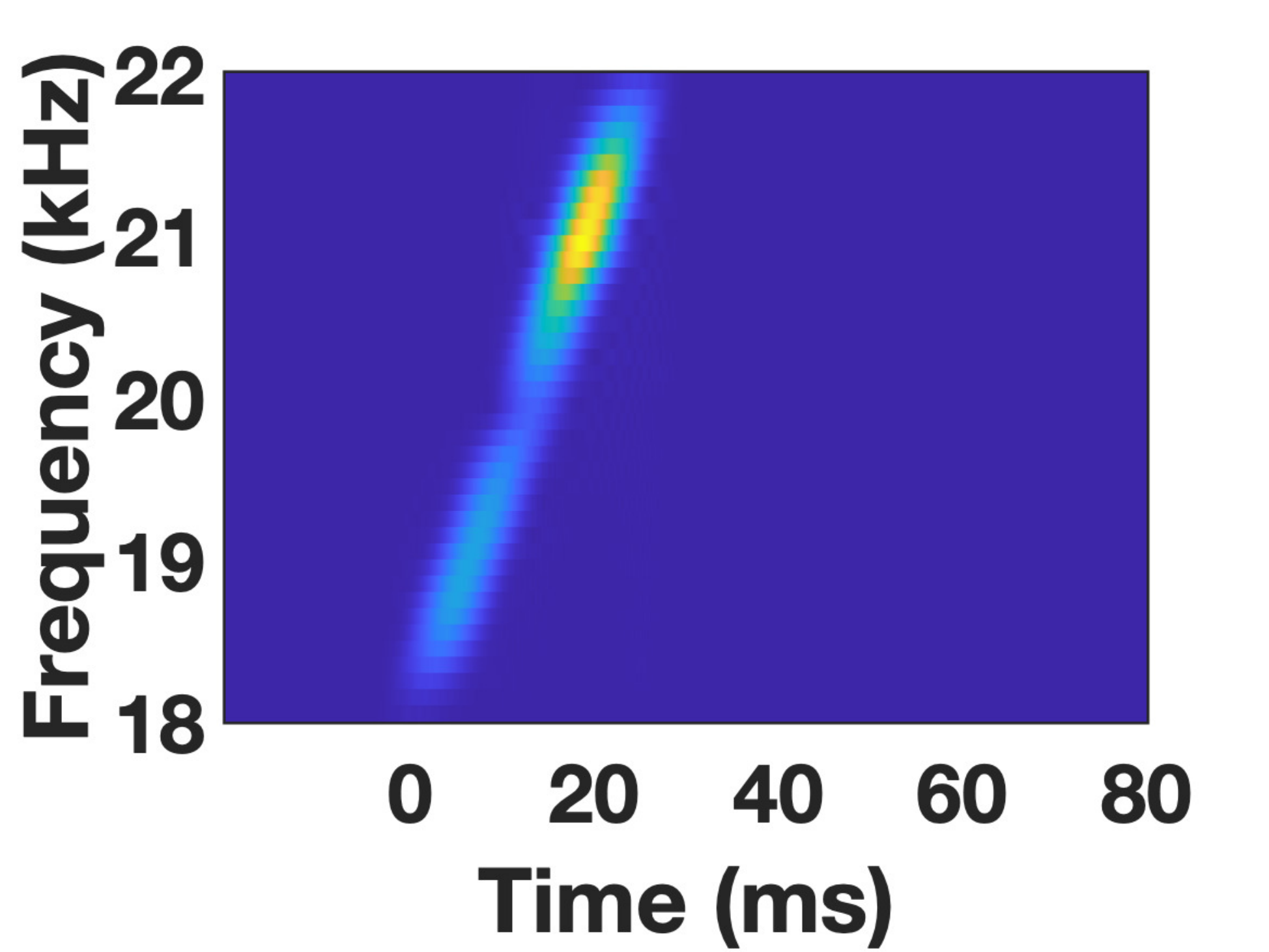}
			&
			\includegraphics[width=0.29\columnwidth]{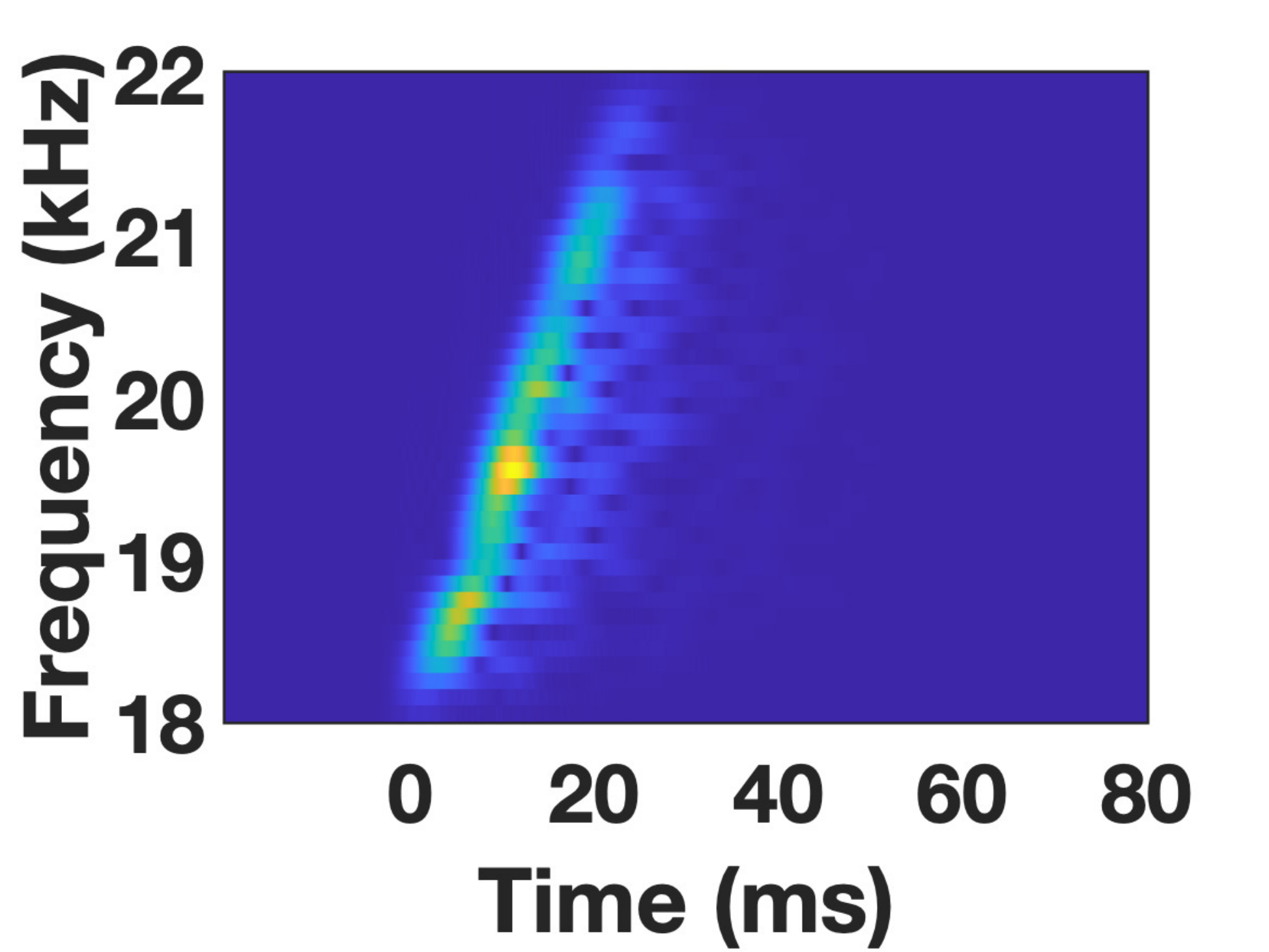}
			&
			\includegraphics[width=0.29\columnwidth]{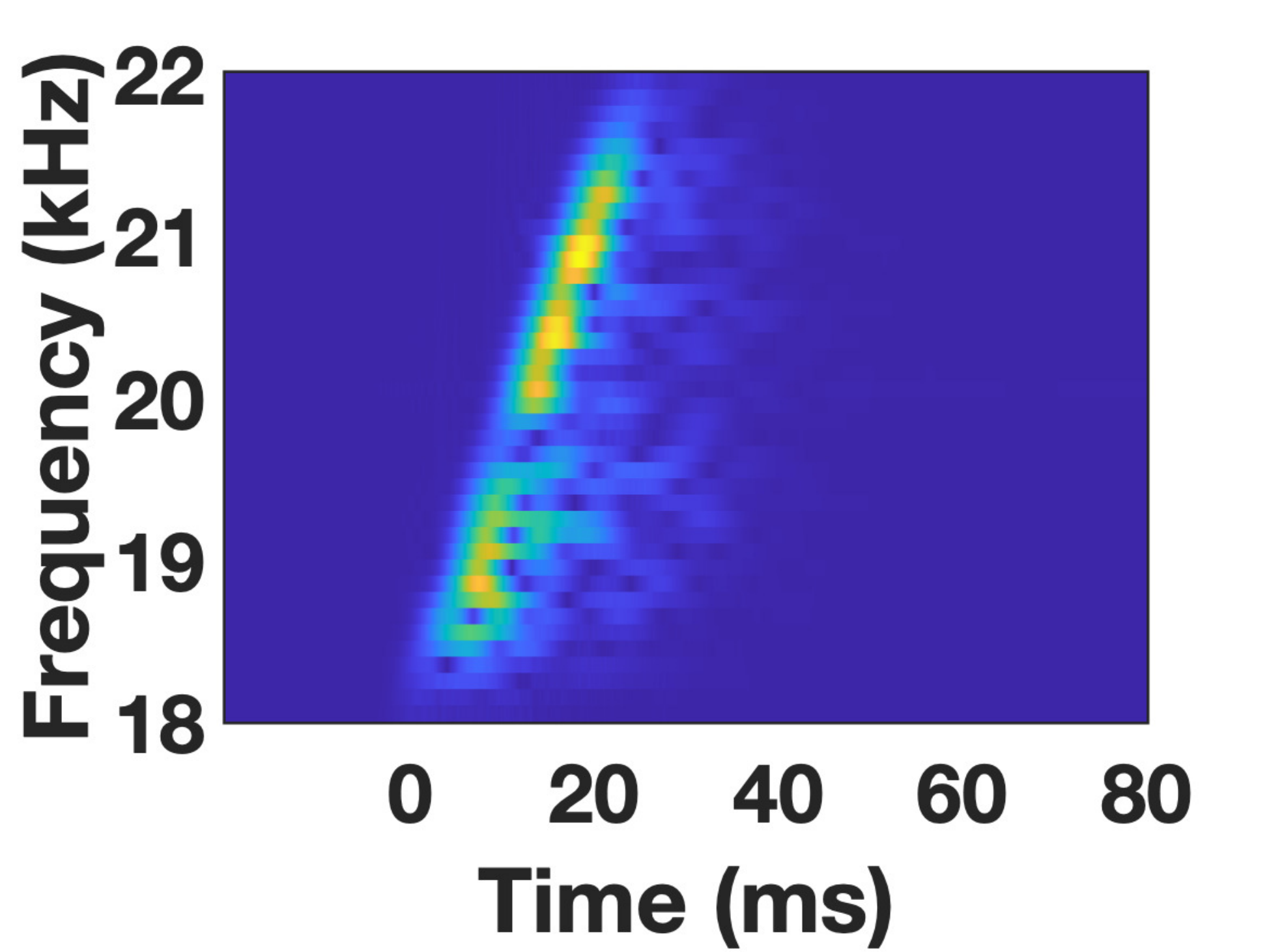}
			\\{\scriptsize(d) In pocket} & {\scriptsize(e) On seat} & {\scriptsize(f) On phone mount} 
		\end{tabular}
	\end{center}
	\vspace{-3mm}
	\caption{Short-time Fourier transform of different phone use statuses. }
	\vspace{-5mm}
	\label{fig:placement_stft}
\end{figure}

We derive the Short-Time Fourier Transform (STFT) from the pulse signal to describe the characteristics of the contact object in the acoustic domain. STFT presents the frequency spectrum along time, which captures how each spectral point of the signal is interfered by the hand or a support surface in the vehicle. In particular, we use a sliding window to compute the Discrete-Time STFT (DT-STFT) of the pulse signal, which results in a 2D image. The value of each image pixel at sample $m$ and frequency $f$ is expressed by Equation \ref{eq:dtstft}, where $w(n)$ is a window function.
\begin{equation}
	\label{eq:dtstft}
	\setlength{\abovedisplayskip}{3pt}
	\setlength{\belowdisplayskip}{3pt}
	DT\mbox{-}STFT(m, f)=\sum_{n=-\infty}^{\infty}\hat x(n)w(n-m)e^{-j2\pi fn}
\end{equation}
Though the derived DT-STFT covers the microphone's all frequencies, which span from 0 to 24kHz, we crop the image to only focus on the pulse signal's frequency range from 18kHz to 22kHz. Figure \ref{fig:placement_stft} shows the feasibility of using the DT-STFT image of the pulse signal to differentiate six different scenarios. We can observe that the DT-STFT images show distinct pixel patterns among all the contact objects. For example, the in-hand scenario presents several strong spectral points around 19kHz, while the center console, pocket and seat show lower amplitudes around this frequency. When the phone is on the center console, cup holder, pocket and phone mount, the received pulse signal has great amplitudes between 20kHz and 22kHz. In comparison, the gripping hand suppresses the pulse signal significantly on these frequencies. The reason is that the impact on the pulse signal depends on the contact object's material, contact area and pressure, which may reinforce the signal at some frequencies but suppress it at the others. Our next step is to use a deep learning algorithm to discriminate the handheld phone use from the other handsfree scenarios based on the DT-STFT images.

\subsection{CNN-based Binary Classifier}
We resize all the DT-STFT images into a fixed size and process them using a binary classifier based on Convolutional Neural Network (CNN). CNN is widely used to analyze images by learning their patterns. To recognize the gripping hand based on DT-STFT images, we develop a CNN-based binary classifier with three convolution layers and one fully connected layer, which is a CNN structure widely used on mobile devices~\cite{xu2019first}. The output dimensions in each layer are tuned to reduce the processing time while ensuring the accuracy. Specifically, the dimensions of the output can be calculated as 
\begin{equation}
	\setlength{\abovedisplayskip}{5pt}
\setlength{\belowdisplayskip}{5pt}
	dimensions = ( \tfrac{m - k + 2d}{l} + 1 ) \times ( \tfrac{m - k + 2d}{l} + 1 ) \times t
\end{equation}
where $m$, $k$, $l$, $d$ and $t$ are the input image size, the kernel size, the step length, the number of padding applied and the number of filters.

\begin{table}[b]
	\centering
	\footnotesize
	\vspace{-5mm}
	\caption{The structure of our CNN-based binary classifier}
	\vspace{-2mm}
	\label{tab:my-table}
	\begin{tabular}{|l|l|l|}
		\cline{1-3}
		\textbf{Layer}                    & \textbf{Output Shape} & \textbf{Param \#}\\ \cline{1-3}
		Input: short-time Fourier transform                & (150, 150, 3)   & 0                 \\ \cline{1-3}
		Conv2D + RecLineU                 & (148, 148, 32)  & 896                \\ \cline{1-3}
		Max Pooling 2D                    & (74, 74, 32)    & 0                  \\ \cline{1-3}
		Conv2D + RecLineU                 & (72, 72, 32)    & 9248               \\ \cline{1-3}
		Max Pooling 2D                    & (36, 36, 32)    & 0                 \\ \cline{1-3}
		Conv2D + RecLineU                 & (34, 34, 32)    & 9248              \\ \cline{1-3}
		Max Pooling 2D                    & (17, 17, 32)    & 0                   \\ \cline{1-3}
		Flatten                           & (9248)          & 0                 \\ \cline{1-3}
		Dropout                           & (9248)          & 0                \\ \cline{1-3}
		Dense                             & (128)           & 1183872           \\ \cline{1-3}
		Dense\_1                          & (60)            & 7740               \\ \cline{1-3}
		Dense\_2                          & (2)             & 122                \\ \cline{1-3}
		Output: Probability in {[}0, 1{]} & (1)             & 0                  \\ \cline{1-3}
	\end{tabular}
\end{table}

The detailed structure of our CNN classifier is as shown in Table \ref{tab:my-table}. In particular, the dimensions of the normalized input image is $150 \times 150$. The convolutional kernel size is $3 \times 3$ and the pooling kernel size is $2 \times 2$. The step length is set as $1$, the number of padding applied is set as $0$, and the number of filters is $32$. The dimensions after the first convolution operation is $148 \times 148 \times 32$ as computed by the above equation. Since the kernel size of the pooling layer is $2 \time 2$, the dimension after the first pooling operation is $74 \times 74 \times 32$. We keep the same configuration for the rest of the convolution and pooling layers. At the end of the model, we utilize the softmax function to normalize the network output and obtain a probability for each class as the decision confidence or CNN score. We then utilize the Adam as the optimizer, which leverages the power of adaptive learning rates to find individual learning rates for each parameter. We use sparse categorical cross-entropy as the model's loss function since we expect class labels to be provided as integers instead of one-hot encodes ones.



Our CNN-based algorithm performs the binary classification to discriminate the handheld and handsfree phone uses, which consists of two phases. During the training phase, we involve a number of people to collect the handheld and handsfree phone-use instances. Moreover, the various handheld phone-use activities are considered in order to cover the various scenarios when the user holds the phone still, tap/swipe on the phone screen and hold the phone close to face (e.g., making phone calls). It is important to note that these phone-use activities generate sounds and cause the handheld status to be unstable. Our system does not rely on these sounds to recognize the handheld phone use, because they differ significantly among different people and different activities. These acoustic noises mainly reside at low frequencies and are suppressed by our bandpass filter. Though the phone can be used differently in the driver's hand, our CNN algorithm can still distinguish them from the handsfree scenarios, as the phone is consistently in the user's hand, which is discernible from other contact objects. Additionally, we separately train two CNN models for the phone's Mic 1 and Mic 2, which analyzes the contact object from two acoustic channels. 

During the testing phase, the DT-STFT images of the testing pulse sound are input to the two CNN models to process independently. The CNN scores of the two models are integrated to make the classification decision. This result is the phone-use status sampled by one sensing pulse.  \textcolor{black}{The running time and memory usage of our model is 150 MFLOPS and 5MB, respectively. Therefore, running our model on mobile processors is practical and inexpensive. Besides, the trained model has a size of 14.8MB, which is portable enough for mobile devices for real-time inference. }

\subsection{Handheld Phone-Use Monitoring} 
The accurate classification of each pulse sound is the basis for monitoring the phone use and detecting distracted driving instances. But monitoring the phone use in practical in-vehicle scenarios is more challenging, as even the classification error of a single sample could come at a tremendous cost. We continue to investigate the practical phone-use monitoring and correct the sample errors to cope with the false positives and false negatives in the classification results. 

Our system is designed to sample the phone-use status ten times per second. The phone-status monitoring result is a sequence of labels between \textit{handheld} and \textit{handsfree}, based on which the system decides when the user grabs or drops off the phone. We design an adaptive window-based error correction filter to process the label sequence based on the \textit{flip-and-merge} rule. The adaptive window starts from the first sample of the current instance and compares it with its adjacent next sample. If their labels are the same, the window grows its size by one and examines the next two consecutive samples. This recursion continues until the sample status changes. The current window extracts a sample chunk, and its size $W$ is recorded. Then, the above process repeats to find the next chunk.

\begin{figure}[t]
	\begin{center}
		\includegraphics[width=3.2in,height=2.4in]{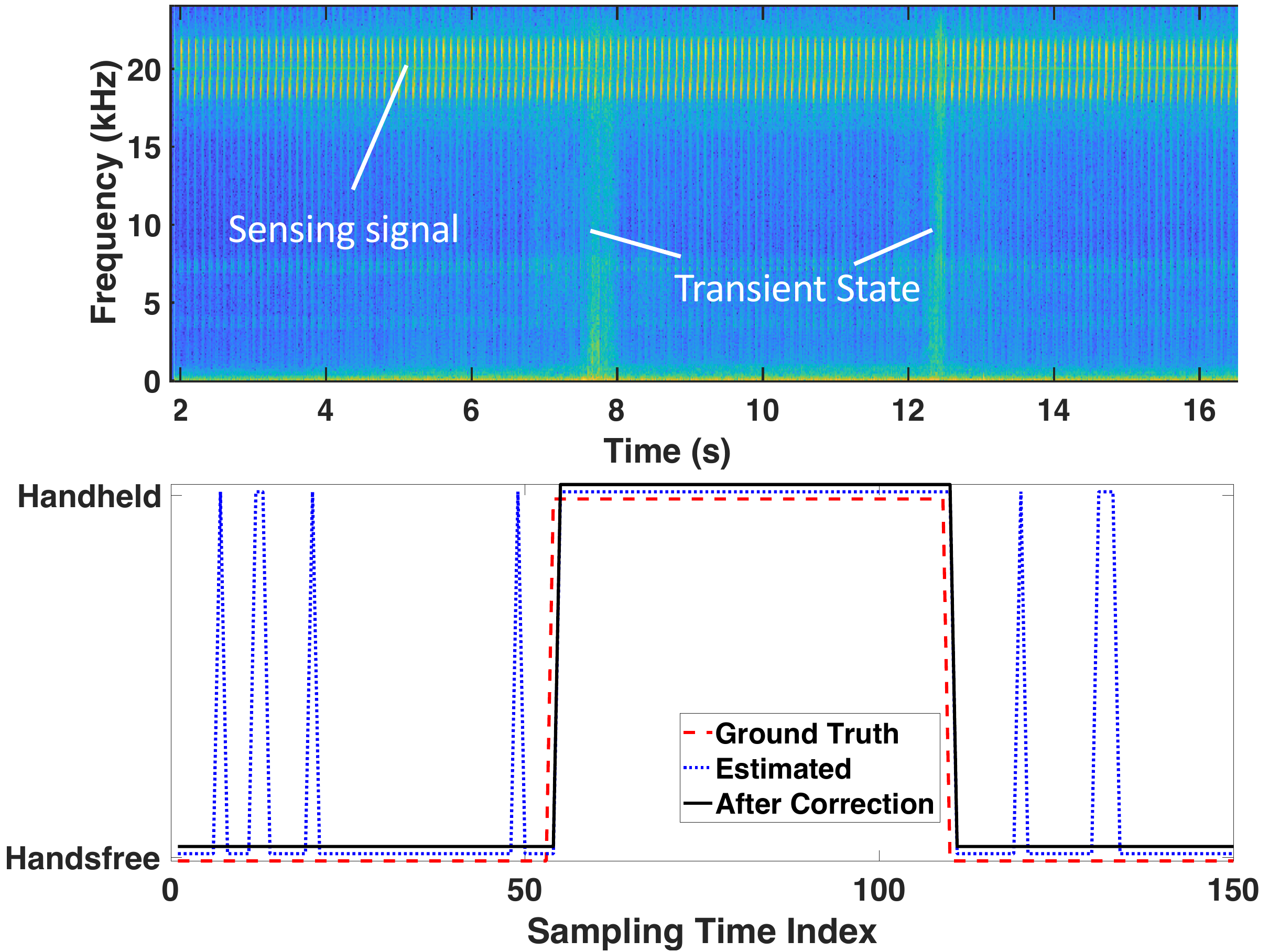}
	\end{center}
	\vspace{-4mm}
	\caption{The phone-use monitoring and classification error correction. }
	\vspace{-5mm}
	\label{fig:error_correction}
\end{figure}

The \textit{flip-and-merge} rule further determines each chunk to be an error or a valid chunk with two thresholds $th_1$, $th_2$ ($th_1 < th_2$), where a valid chunk represents a complete or a partial instance. The intuition is that when a driver uses a phone, the duration can not be too short (even for checking time). 
If $W \geq th_2$, the chunk is determined to be a valid chunk. If $W < th_1$, the entire chunk is considered to be misclassified because the phone status toggles back and forth too fast, and the labels of its all samples are flipped. This chunk after correction is merged to its closest valid chunk. If $th_1 < W < th_2$, we need to examine the labels of its two valid neighbor chunks, $v_{pre}$ and $v_{next}$. If $v_{pre} = v_{next}$, we consider this current chunk to be erroneous, so that it is flipped and merged with its neighbor chunks. If $v_{pre} \neq v_{next}$, we keep the label of the current chunk and merge it with the valid neighbor chunk that has the same label. As a result, the handsfree and handheld instances are obtained. Especially, the handheld instance is detected, if the prior chunk is a handsfree instance and the current chunk size grows larger than $th_2$ (it is not necessary to wait for obtaining an entire chunk). The first sample of the current chunk then captures the handheld instance start, and its end is determined when the next chunk is confirmed to be a handsfree instance. Empirically, we use $0.5$s and $0.8$s for $th_1$ and $th_2$.

Figure~\ref{fig:error_correction} illustrates the phone-use monitoring when a driver grabs the phone for 5 seconds and then drops it off. The top figure presents the spectrogram of this process, where the ultrasonic pulses periodically sense the phone-use status, and the transient state sounds (i.e., phone-grab and drop-off actions) show the main signal powers at lower frequencies. The bottom figure illustrates the phone-use status monitoring results. We observe that though some samples are mistakenly classified, they can be corrected by our adaptive window-based filter. The resulted phone-use status sequence is close to the ground truth curve. From this monitoring result, we can detect the complete distracted driving instance as well as determining its start, end and duration.

	  \section{Performance Evaluation}
\begin{figure}[t]
	\begin{center}
		\includegraphics[width=3.0in]{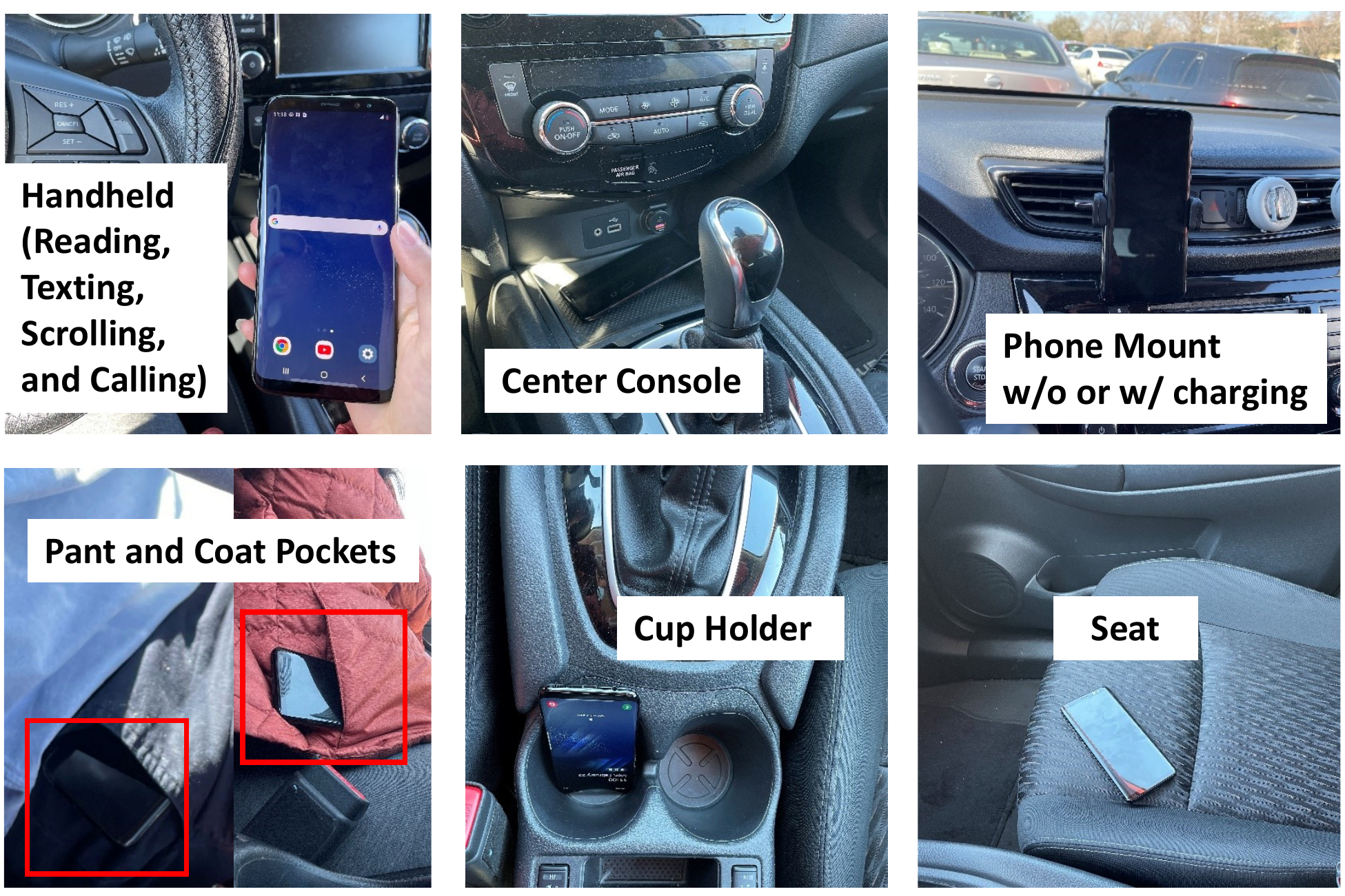}
	\end{center}
	\vspace{-2mm}
	\caption{Eleven experimental scenarios in the vehicle. }
		\vspace{-5.4mm}
	\label{fig:setup}
\end{figure}

\subsection{Experimental Setup}
To evaluate our system, we develop an experimental platform based on Android, which periodically sends ultrasonic pulse signals and records the stereo sounds simultaneously. We use this platform to collect data from three smartphone models, Samsung Galaxy S8, Motorola Moto G8, and Google Pixel2, and the data is processed offline. These devices run Android 9.0, and the microphone sampling rate is set to 48kHz. We also test two vehicle models, Nissan Rogue (Car A) and Volkswagen Tiguan (Car B). \textcolor{black}{We develop our CNN-based binary classifier based on Keras 2.4.3.} We recruited fourteen participants for data collection. As shown in Figure~\ref{fig:setup}, each participant was asked to use the phone in \textcolor{black}{eleven scenarios}, including four handheld phone uses (i.e., holding the phone still or reading, texting, scrolling and calling) and seven handsfree scenarios (i.e., in a coat pocket, pant pocket, cup holder, center console, phone mount, \textcolor{black}{phone charging on phone mount} and seat). For each scenario, the participant was asked to re-grab or reposition the phone 40 times to include behavioral inconsistency and phone location differences. We apply half data for training and half for testing. 

The overall performance is evaluated based on \textcolor{black}{fourteen participants, eleven scenarios,} car A and Samsung S8 with city driving. We also investigate the various impact factors based on four participants and  \textcolor{black}{eleven} scenarios. In particular, the device model and the car model impacts are studied. Moreover, four typical in-vehicle environments \textit{engine off}, \textit{engine on}, \textit{city driving} and \textit{highway driving} are tested and compared, where the practical in-vehicle noises caused by the engine, road conditions and traffic are involved. Additionally, the impact of the car audio (e.g., radio sounds) is studied. Furthermore, we monitored four participants' phone use while driving on the highway, in which \textcolor{black}{each participant was asked to use the phone by grabbing it 40 times} from the seat, center console, cup holder, phont mount and pocket for an hour monitoring. Due to safety reasons, when the car was moving, the phone-use experiments were performed by the front passenger. 

\subsection{Overall Performance }

\begin{figure}[t]
	\begin{center}
		\includegraphics[width=2.5in]{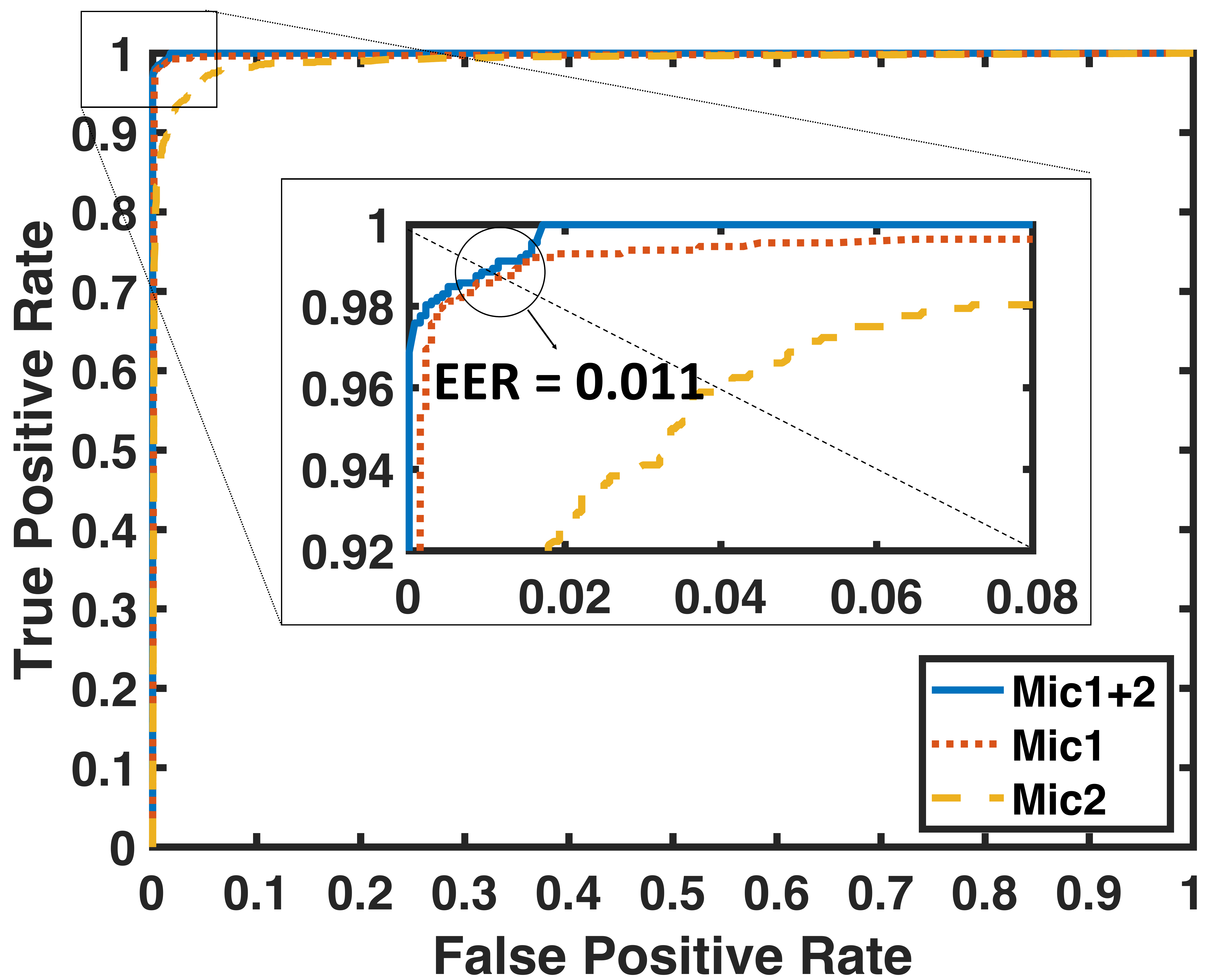}
	\end{center}
	\vspace{-3mm}
	\caption{Overall performance of our system. }
		\vspace{-4.2mm}
	\label{fig:overall_roc}
\end{figure}

\subsubsection{Handheld vs. Handsfree}
The ROC curves of our system to recognize the phone-use status are presented in Figure \ref{fig:overall_roc}. We find the system achieves a high True Positive Rate (TPR) and low False Positive Rate (FPR) to discriminate \textit{handheld} from \textit{handsfree}. In particular, when integrating the two microphones, \textcolor{black}{our system achieves 98.4\% TPR and 0.5\% FPR, and the Equal Error Rate (EER) is 1.1\%.} The result is very promising as the system correctly recognizes the handheld and handsfree scenarios, regardless of how the driver uses the phone and who holds the phone. The result also indicates that our system is effective in practical city driving scenarios. Furthermore, we find Mic 1 performs better than Mic 2. The reason is that Mic 1 is at the top of the phone, which is far from the bottom speaker. Compared to Mic 2 that is close to the speaker, the Mic 1 received sounds travel across the phone case and interact better with the contact object to capture its characteristics. 

\subsubsection{Phone-use Scenarios}
\begin{figure}[t]
	\begin{center}
		\includegraphics[width=0.8\columnwidth,height=1.4in]{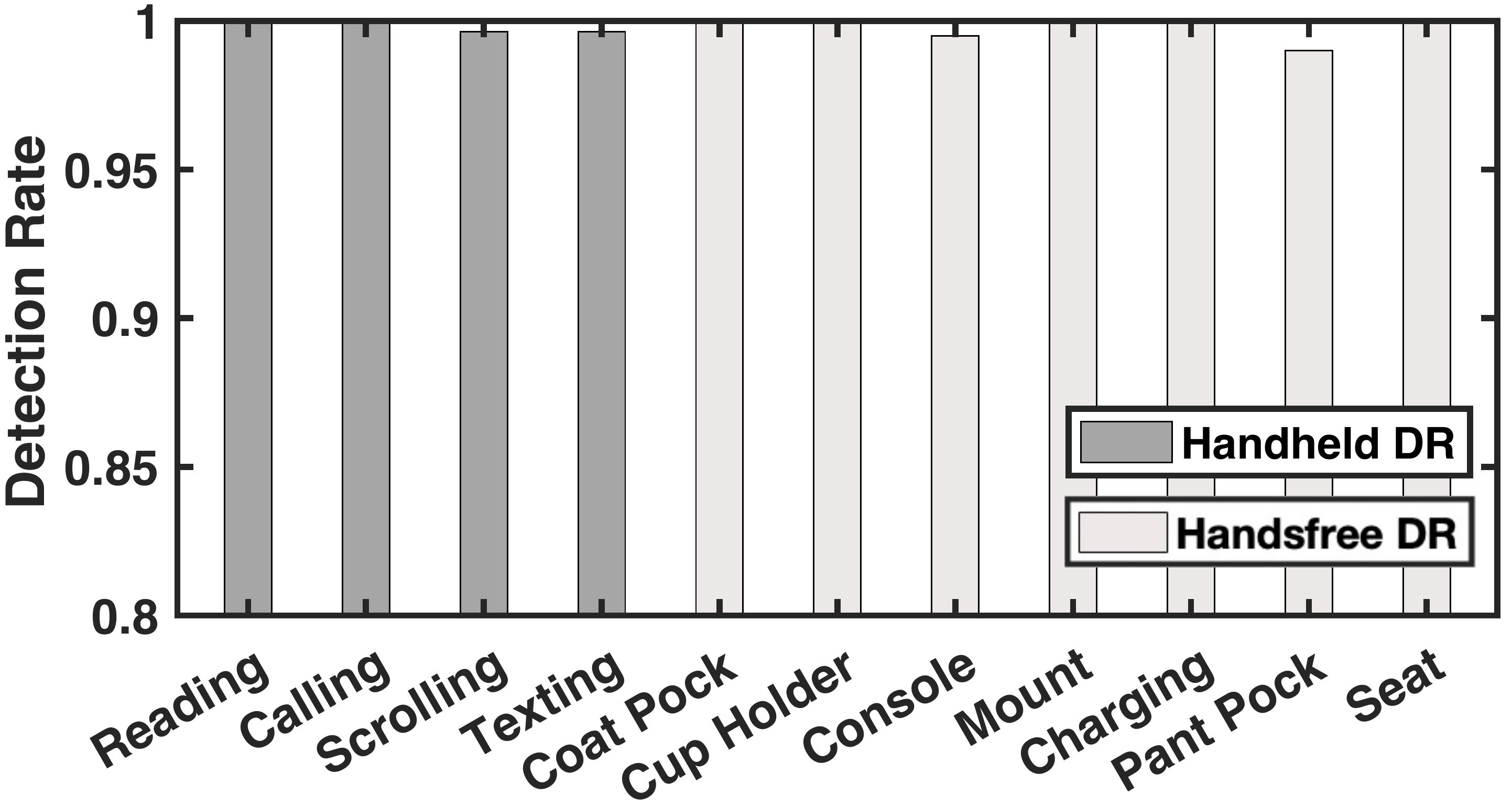}
	\end{center}
	\vspace{-4mm}
	\caption{Performance under different phone-use contexts. }
	\vspace{-3mm}
	\label{fig:overall_by_scenario}

\end{figure}

Next, we investigate how the system discriminates each of the eleven phone statuses between \textit{handheld} and \textit{handsfree}. 
Figure \ref{fig:overall_by_scenario} presents \textcolor{black}{gripping hand Detection Rate (DR) in four handheld scenarios and handsfree DR in seven handsfree scenarios}. We observe that our system performs well for all eleven scenarios, obtaining a mean \textcolor{black}{99.8\% DR.} For example, \textcolor{black}{reading and calling} perform the best among the four handheld scenarios with a 100\% accuracy. \textcolor{black}{The DRs of scrolling and texting are slightly lower, which are 99.6\%. The reason is that the hand movements in the two scenarios cause noises and the slightly unstable contact relationship between the phone and the hand.  
For the seven handsfree scenarios, except for console and pant pocket that perform with 99.5\% DR and 99.0\% DR, respectively, the other five scenarios all achieve a 100\% DR to be recognized as handsfree.} The results indicate that our system successfully classifies the phone-use scenarios based on their contact with the phone. 

\subsubsection{Individual Difference}
\begin{figure}[t]
	\begin{center}
		\includegraphics[width=0.8\columnwidth,height=1.4in]{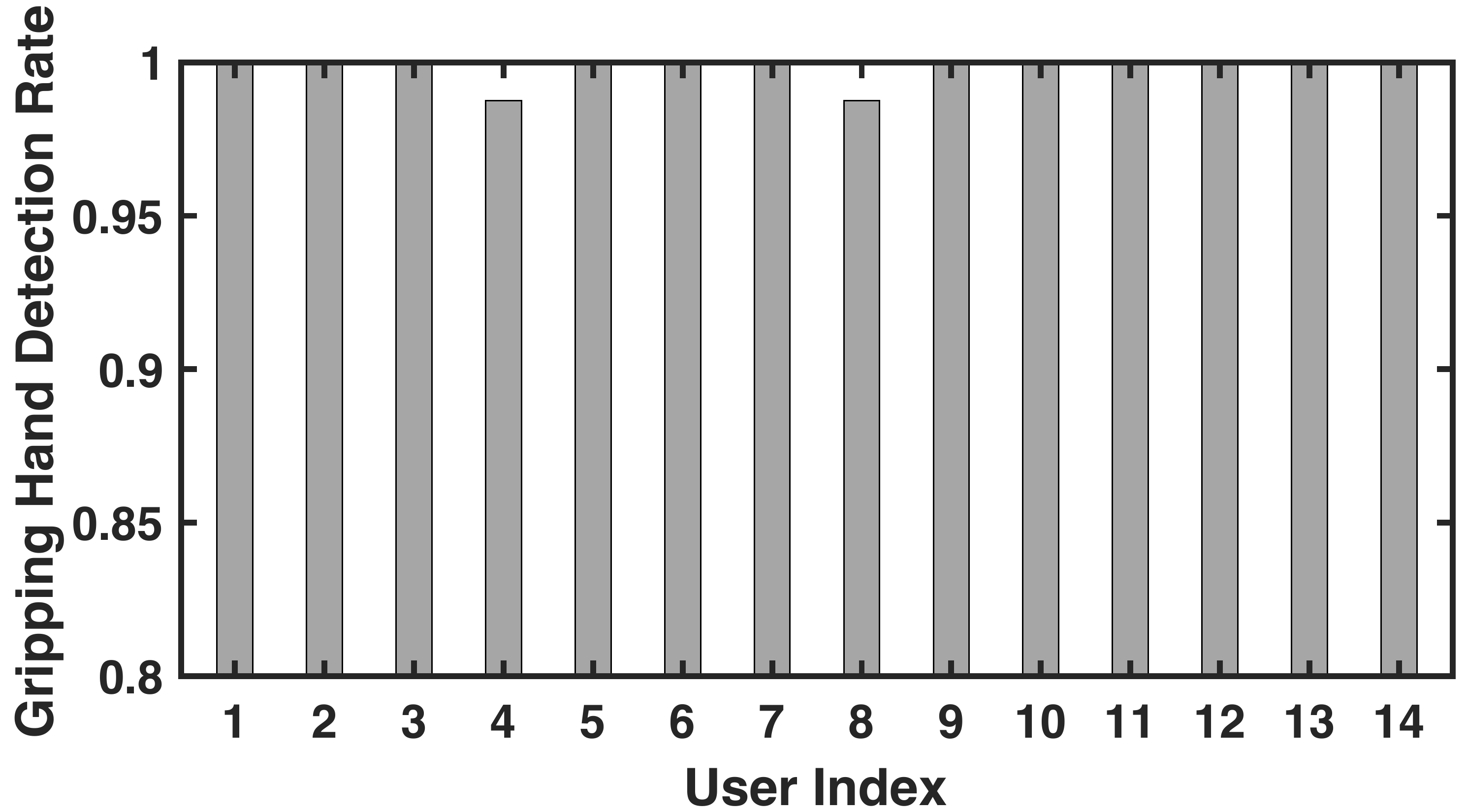}
	\end{center}
	\vspace{-4mm}
	\caption{Handheld device distraction detection for different users. }
	\vspace{-4mm}
	\label{fig:overall_by_user}
\end{figure}

We also study how the system performs for different users. Figure \ref{fig:overall_by_user} presents the DR of four types of handheld phone-use activities for \textcolor{black}{fourteen users}. We observe that the system accurately detects the handheld phone use for all participants with an average \textcolor{black}{99.8\% DR.} Moreover, more than half of the users achieve a DR of 100\%, and the lowest DR is \textcolor{black}{98.7\%.} The results show that our system can work for different users regardless of their unique hand geometry and gripping strengths.

\subsection{Impact Factors }
\subsubsection{Device Models}

\begin{figure}[t]
	\begin{minipage}{.49\columnwidth}
		\includegraphics[width=1.6in]{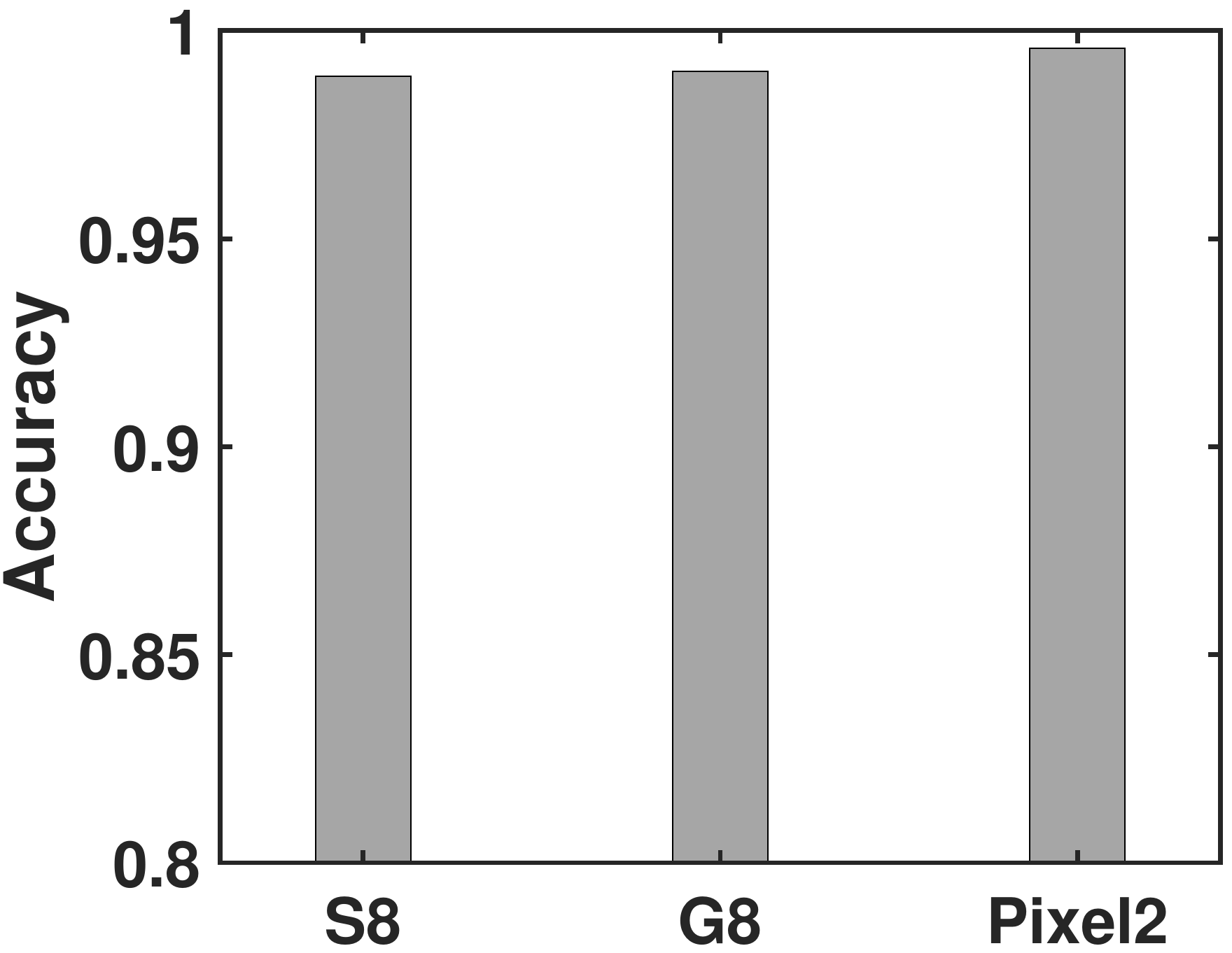}
		\caption{Impact of different devices. }
		\label{fig:accuracy_by_device}
	\end{minipage}
	\begin{minipage}{.49\columnwidth}
		\includegraphics[width=1.6in]{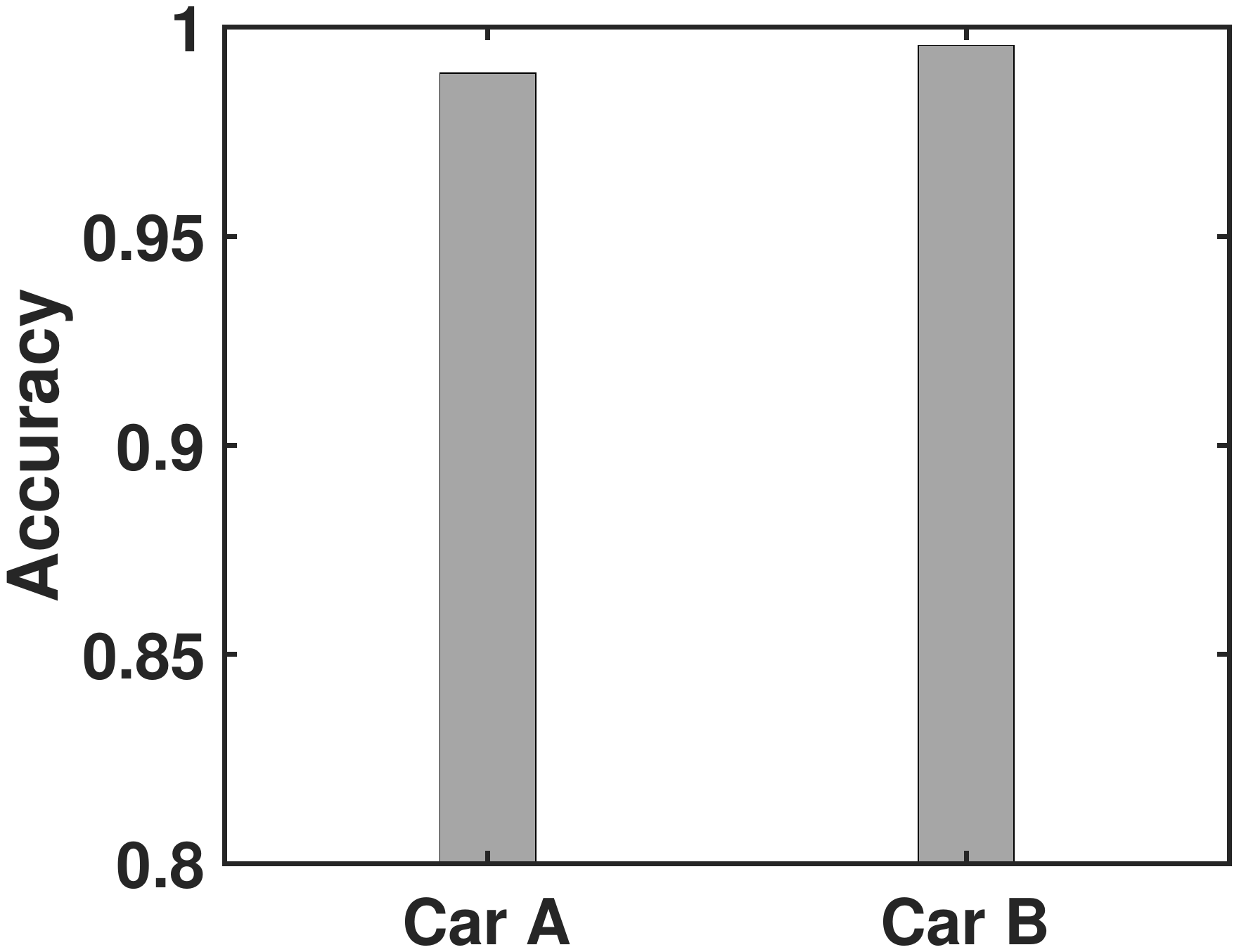}
		\caption{Impact of car models. }
		\label{fig:accuracy_by_car}
	\end{minipage}
\end{figure}

We now investigate the impacts of device models. Our participants were asked to use three different phones in Car A, and the above eleven types of phone statuses were collected during city driving. 
Figure \ref{fig:accuracy_by_device} shows the classification accuracy for each device. We observe that all three devices accurately distinguish the handheld phone uses from the handsfree. In particular, Google Pixel 2 performs the best with 99.6\% accuracy. The performances of Samsung Galaxy S8 and Motorola G8 are slightly lower, which are at 98.9\% and 99.0\%, respectively. The results indicate that our system can be broadly deployed on different devices.

\subsubsection{Car Models}
Similarly, the shells and interiors of different car models may affect the performance of our system. Therefore, we repeat the above experiments in Car B using Samsung Galaxy S8. Figure \ref{fig:accuracy_by_car} shows the performance of each car model. It can be observed that both car models achieve a good performance. In particular, Car B achieves an accuracy of 99.6\%, which is slightly higher than Car A. The reason may be that Car B has a thick shell, which suffers less from the wind, road and engine noises. The results show our system is able to detect distracted driving with different car models.

\subsubsection{Vehicle Engine Status} 
\begin{figure}[t]
	\begin{minipage}{.49\columnwidth}
		\includegraphics[width=1.6in]{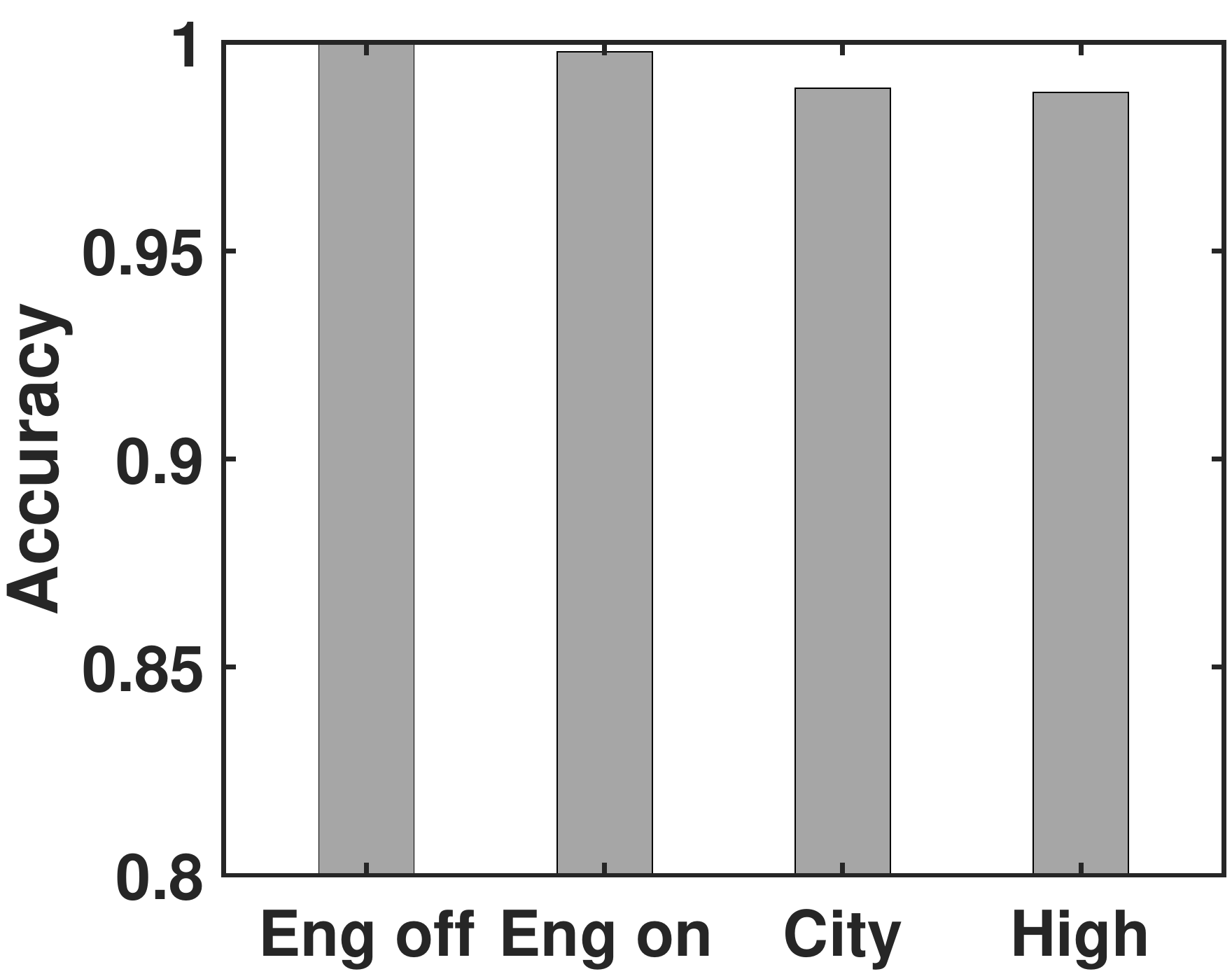}
		\caption{Impact of engine status. }
		\label{fig:accuracy_by_engine_status}
	\end{minipage}
	\begin{minipage}{.49\columnwidth}
		\includegraphics[width=1.6in]{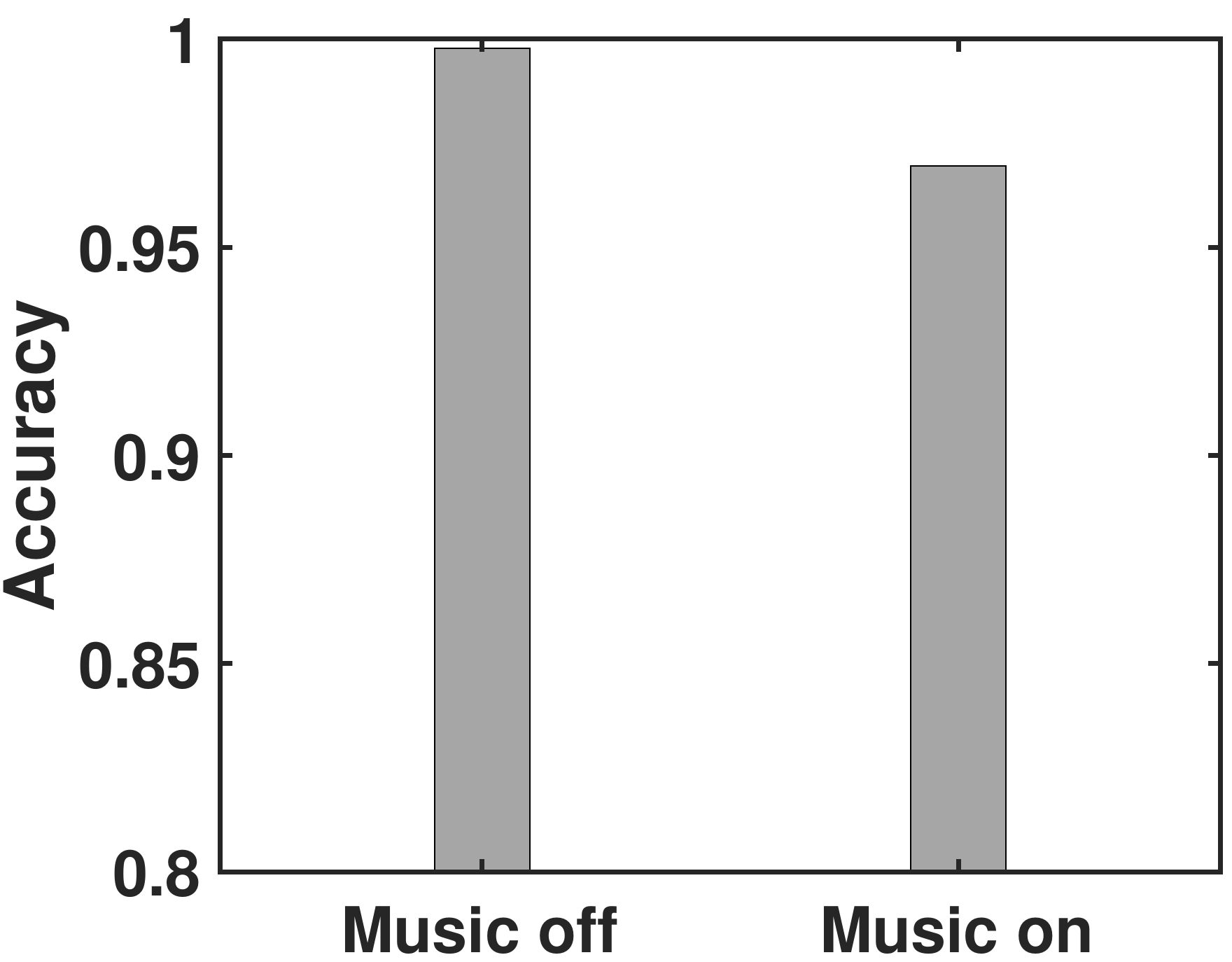}
		\caption{Impact of in-car music. }
		\label{fig:accuracy_by_noise}
	\end{minipage}
		\vspace{-4mm}
\end{figure}

The car engine at different statuses or speeds generates different levels of noises, including increased or decreased road and wind noises. We thus evaluate our system under different engine statuses, including \textit{city driving}, \textit{highway driving}, \textit{engine on} and \textit{engine off}. We use Car A and Samsung Galaxy S8 for this impact study. Figure \ref{fig:accuracy_by_engine_status} presents the classification results under the four different engine status. Not surprisingly, \textit{engine off} performs the best with 100\% accuracy, as this is a quiet in-vehicle environment. \textit{Engine on} also performs well with 99.8\% accuracy. \textit{City driving} and \textit{highway driving} achieve a slightly lower accuracy, which are 98.9\% and 98.8\% respectively, though they suffer from different types of noises. In particular, \textit{city driving} mostly involves the noise from frequent accelerations and braking in the traffic, while \textit{highway driving} experience more engine and wind noises. However, our system is robust enough to detect the phone-use distraction in both driving environments.

\subsubsection{Car Audios}
When driving, the drivers may turn the radio or music on. The car audio sounds may interfere our sensing signal and affect the system performance. We therefore evaluate our system with the car music on. The experiment was done with Car A and Samsung Galaxy S8, under the \textit{engine on} status. The music sounds were between $56\sim60dB$. Figure \ref{fig:accuracy_by_noise} compares the performances of our system when the music is on or off. We observe that the music sounds do have a slight impact on our system performance. The classification accuracy degrades to 97.0\% when the music is on, which is still high. The result confirms the robustness of our system to work under car audios.

\subsection{Phone-use Monitoring Case Study}

\begin{figure}[t]
	\begin{minipage}{.49\columnwidth}
		\includegraphics[width=1.7in]{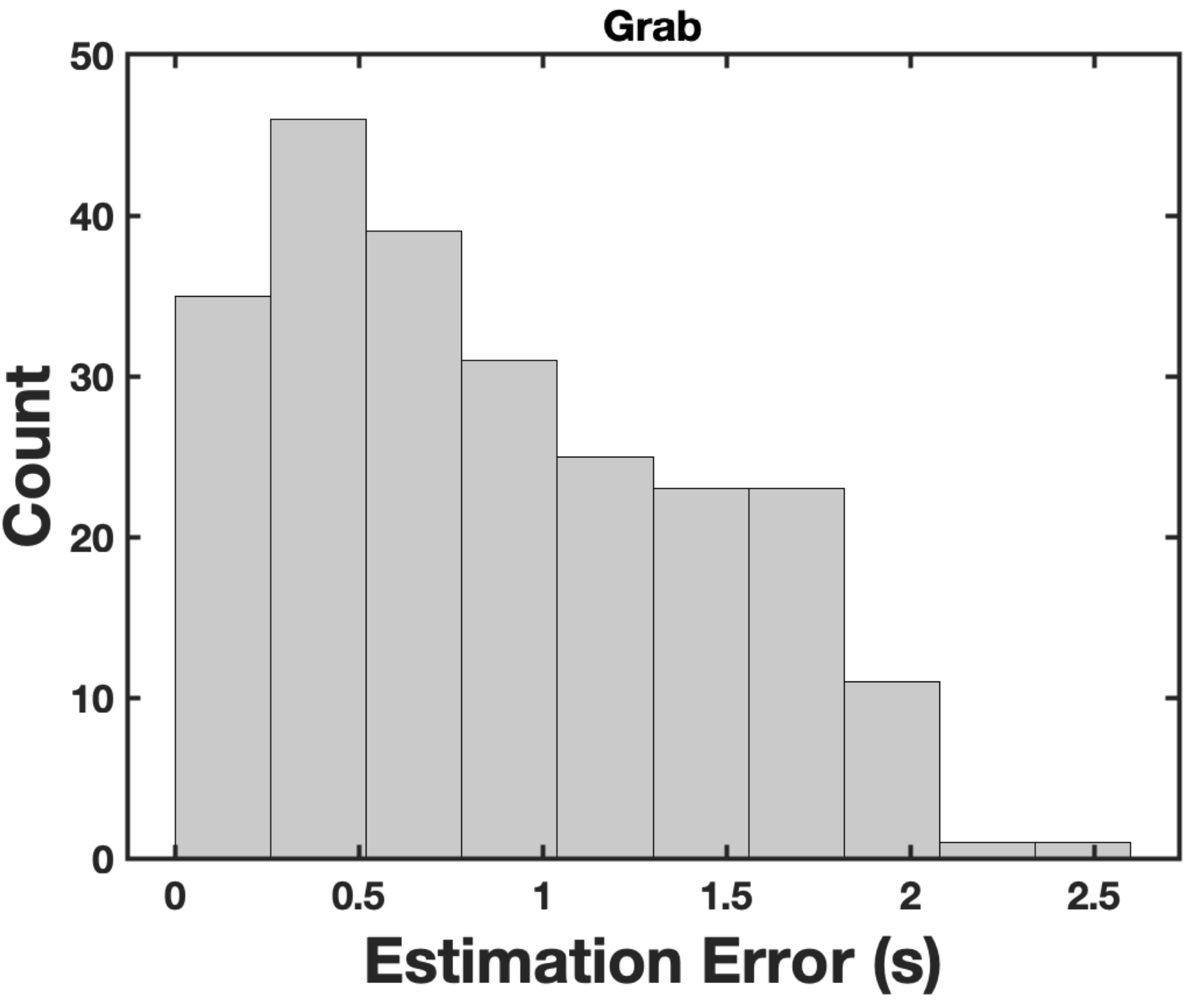}
		\caption{Distraction start estimation. }
		\label{fig:estimation_error_grab}
	\end{minipage}
	\begin{minipage}{.49\columnwidth}
		\includegraphics[width=1.7in]{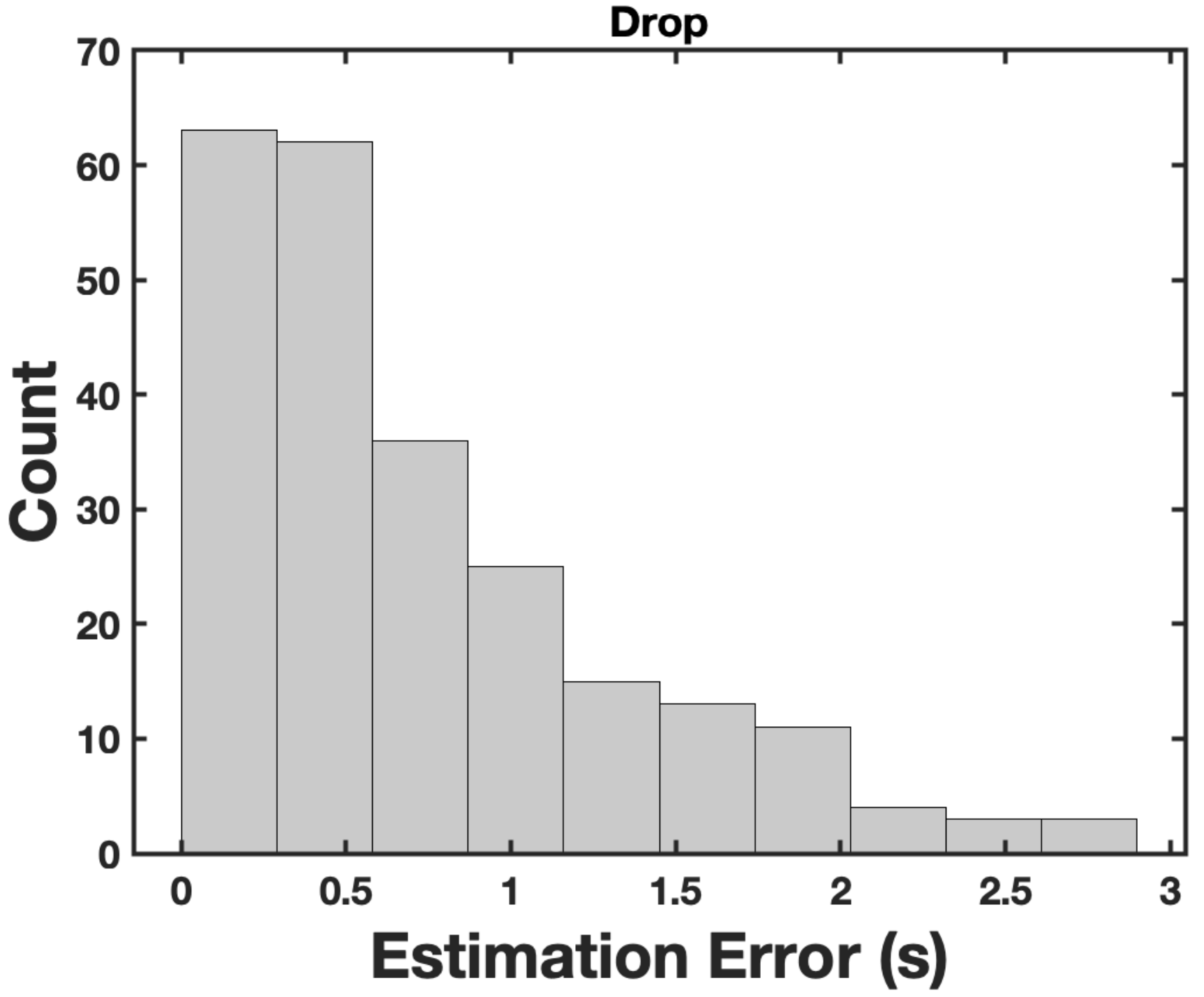}
		\caption{Distraction end estimation. }
		\label{fig:estimation_error_drop}
	\end{minipage}
		\vspace{-4mm}
\end{figure}

Lastly, we conducted a case study to monitor four participants' phone use during highway driving and detect their distracted driving activities. Our system achieves \textcolor{black}{99.5\%} detection rate to capture all the distracted driving activities of these participants, and the false positive rate is \textcolor{black}{0.4\%.} The promising performance is the result of both high classification performance and the system's error correction capability.  Figure~\ref{fig:estimation_error_grab} and Figure~\ref{fig:estimation_error_drop} further present the distributions of the absolute time errors to determine the start and end of each distracted activity. Our system achieves a median error of \textcolor{black}{0.76 second} to determine the start of a distracted driving activity. The median error to determine the end time is \textcolor{black}{0.55 second.} These errors are mainly resulting from the transient states when the user grabs and drops the phone. \textcolor{black}{We also find that most of the larger errors (e.g., between 1s and 2s) occur when the user grabs the phone from or drops it to a phone mount, because these actions are relatively less smooth resulting a longer transient time. }
The results confirm that our system is effective to monitor the driver's phone use. 

	\vspace{-2mm}
\section{Conclusion}
\vspace{-1mm}
This work proposes a phone-use monitoring system to address handheld phone distractions by sensing the user's gripping hand. By emitting periodic ultrasonic pulse signals for sensing, the proposed system continuously discriminates whether the phone is held by hand or placed on a support surface in the vehicle, such as seat, center console, pocket and phone mount. When detecting that the driver's hand reaches the phone, the system disables all phone services except emergency calls to eliminate the distraction. In particular, we derive the short-time Fourier transform of each received pulse signal to capture the unique impacts imposed by the gripping hand and the various support surfaces. Moreover, we develop a CNN-based binary classifier to analyze the STFT images and discriminate the phone use between the handheld and the handsfree status. We further design an adaptive window-based filter to correct the classification errors to capture each complete handheld phone-use activity, as well as determining its start, end and duration. Extensive experiments in practical driving environments show that our system detects the smartphone handheld status with 99\% accuracy and captures the start of the distraction period with \textcolor{black}{0.76 second} median error. 


	
	\vspace{0.1mm}
	\textbf{Acknowledgment.} This work was partially supported by LEQSF(2020-23)-RD-A-11.
	
	\bibliographystyle{IEEEtran}
	\bibliography{../bib/bib}  
	

\end{document}